\documentclass{article}
\usepackage[boxed,ruled,vlined,linesnumbered,noresetcount]{algorithm2e}
\usepackage{amsmath,latexsym,amsopn,amsfonts,amssymb,amsthm} 
\usepackage{soul}
\usepackage{enumitem}
\usepackage{graphicx,xspace}
\usepackage[utf8]{inputenc}

\newtheorem{theorem}{Theorem}[section]
\newtheorem{lemma}[theorem]{Lemma}
\newtheorem{definition}{Definition}[section]

\newtheorem{assumption}[theorem]{Assumption}

\newcommand\bull{{\operatorname{-\xspace}}}
\newcommand{\blitza}{\text{\usefont{U}{ulsy}{m}{n}\symbol{'011}}}
\newcommand{\mdReset}{{\mathsf{invoc}}\xspace} 
\newcommand{\mdCnt}{{\mathsf{rtComp}}\xspace} 
\newcommand{\trusted}{{\mathsf{trusted}}\xspace}

\usepackage{color}
\newcommand{\emsO}[1]{\textcolor{black}{#1}}
\newcommand{\algSize}{small} 
\newcommand{\bigO}{\mathcal{O}\xspace}
\newcommand{\remove}[1]{}
\newcommand{\reduce}[1]{} 

\newcommand{\Correct}{\mathit{Correct}\xspace} 
\newcommand{\init}{\texttt{init}\xspace} 
\newcommand{\echo}{\texttt{echo}\xspace} 
\newcommand{\ready}{\texttt{ready}\xspace}

\newcommand{\valid}{\texttt{valid}\xspace} 
\newcommand{\etal}{\emph{et al.}\xspace}
\newcommand{\eg}{\emph{e.g.,}\xspace}

\newcommand{\ie}{\emph{i.e.,}\xspace}
\newcommand{\Ie}{\emph{I.e.,}\xspace}

\newcommand{\true}{\mathsf{True}\xspace}

\newcommand{\false}{\mathsf{False}\xspace}

\newcommand{\sP}{\mathcal{P}\xspace}
\newcommand{\bZ}{{Z}\xspace} 

\newcommand{\capacity}{\mathsf{channelCapacity}\xspace} 
\newcommand{\done}{\mathsf{result}\xspace}
\newcommand{\bcdone}{\mathsf{result}\xspace}

\newcommand{\BB}{\vspace*{-\medskipamount}}

\newcommand{\F}{\vspace*{\smallskipamount}}

\newenvironment{claimProof}{\par\noindent\textbf{Proof of Claim  \clmcnt\space}}{\hfill $\Box_{Claim ~ \clmcnt}$}

\newenvironment{lemmaProof}{\par\noindent\textbf{Proof of Lemma  \lemcnt\space}}{\hfill $\Box_{Lemma ~ \lemcnt}$}

\newenvironment{theoremProof}{\par\noindent\textbf{Proof of Theorem  \thmcnt\space}}{\hfill $\Box_{Theorem ~ \thmcnt}$}
\newcommand{\clmcnt}{0}
\newcommand{\lemcnt}{0}
\newcommand{\thmcnt}{0}

\newcommand{\Section}[1]{\section{#1}}
\newcommand{\Subsection}[1]{\subsection{#1}}
\newcommand{\Subsubsection}[1]{\subsubsection{#1}}

\newcommand{\BSubsubsubsection}[1]{\F \noindent $\bullet$ \emph{#1}~~~~~~}

\usepackage{wrapfig}
\usepackage{url,fullpage}
\begin{document}
\title{Self-stabilizing Byzantine Fault-tolerant Repeated Reliable Broadcast}
	
\author{Romaric Duvignau~\footnote{Chalmers University of Technology, Sweden \texttt{\{duvignau,elad@\}chalmers.se}} \and Michel Raynal~\footnote{Institut Universitaire de France IRISA, France \texttt{michel.raynal@irisa.fr}}  \and Elad Michael Schiller~{$^\ast$}}
		
%

%
%
%
%

 \maketitle	
	
\begin{abstract}
We study a well-known communication abstraction called Byzantine Reliable Broadcast (BRB). This abstraction is central in the design and implementation of fault-tolerant distributed systems, as many fault-tolerant distributed applications require communication with provable guarantees on message deliveries. Our study focuses on fault-tolerant implementations for message-passing systems that are prone to process-failures, such as crashes and malicious behavior. 
At PODC 1983, Bracha and Toueg, in short, BT, solved the BRB problem. BT has optimal resilience since it can deal with $t < n/3$ Byzantine processes, where $n$ is the number of processes. The present work aims at the design of an even more robust solution than BT by expanding its fault-model with self-stabilization, a vigorous notion of fault-tolerance. In addition to tolerating Byzantine and communication failures, self-stabilizing systems can recover after the occurrence of \emph{arbitrary transient-faults}. These faults represent any violation of the assumptions according to which the system was designed to operate (provided that the algorithm code remains intact).
We propose, to the best of our knowledge, the first self-stabilizing Byzantine fault-tolerant (BFT) solution for repeated BRB in signature-free message-passing systems (that follows BT's problem specifications). Our contribution includes a self-stabilizing variation on a BT that solves a single-instance BRB for asynchronous systems. We also consider the problem of recycling instances of single-instance BRB. Our self-stabilizing BFT recycling for time-free systems facilitates the concurrent handling of a predefined number of BRB invocations and, by this way, can serve as the basis for self-stabilizing BFT consensus.   
\end{abstract}


\Section{Introduction}
\label{sec:intro}
%
%
%
Fault-tolerant distributed systems are known to be hard to design and verify. High-level communication primitives can facilitate such complex challenges. These high-level primitives can be based on low-level ones, such as the one that allows processes to send a message to only one other process at a time. Hence, when an algorithm wishes to broadcast message $m$ to all processes, it can send $m$ individually to every other process. Note that if the sender fails during this broadcast, it can be the case that only some of the processes have received $m$. Even in the presence of network-level support for broadcasting or multicasting, failures can cause similar inconsistencies. In order to simplify the design of fault-tolerant distributed algorithms, such inconsistencies need to be avoided. Many examples show how fault-tolerant broadcasts can significantly simplify the development of fault-tolerant distributed systems, \eg  State Machine Replication~\cite{DBLP:books/sp/Raynal18} and Set-Constrained Delivery Broadcast~\cite{DBLP:conf/opodis/AuvolatRT19}. The weakest variant, named \emph{Reliable Broadcast}, lets all non-failing processes agree on the set of delivered messages. This set includes all the messages broadcast by the non-failing processes. 
Stronger reliable broadcasts variants specify additional requirements on the delivery order. Such requirements can simplify the design of fault-tolerant distributed consensus, which allows reaching, despite failures, a common decision based on distributed inputs. Reliable broadcast and consensus (as well as message-passing emulation of read/write registers~\cite{DBLP:books/sp/Raynal18}) are closely related to distributed computing problems. This work aims to design an reliable broadcast solution that is more fault-tolerant than the state of the art.

\Subsection{The problem}
\label{sec:problem}    
%
%
Lamport, Shostak, and Pease~\cite{DBLP:journals/toplas/LamportSP82} said that a process commits a Byzantine failure if it deviates from the algorithm instructions, say, by deferring (or omitting) messages that were sent by the algorithm or sending fake messages. Such malicious behavior can be the result of hardware malfunctions or software errors as well as coordinated malware attacks. 
Bracha and Toueg~\cite{DBLP:journals/jacm/BrachaT85,DBLP:conf/podc/BrachaT83}, BT from now on, proposed the communication abstraction of Byzantine Reliable Broadcast (BRB), which allows every process to invoke the $\mathsf{brbBroadcast}(v)$ operation and raise the $\mathsf{brbDeliver}()$ event upon message arrival. Following Raynal~\cite[Ch. 4]{DBLP:books/sp/Raynal18}, we consider the (single instance) BRB problem.
%
%


\Subsubsection{\emsO{Single-instance BRB.}} 
%
\emsO{We require $\mathsf{brbBroadcast}(v)$ and $\mathsf{brbDeliver}()$ to satisfy Definition~\ref{def:prbDef}.}
\begin{definition}
	\label{def:prbDef}
	\begin{itemize}[topsep=0.2em, partopsep=0pt, parsep=0pt, itemsep=0pt,leftmargin=0pt, rightmargin=0pt, listparindent=0pt, labelwidth=0.15em, labelsep=0.35em, itemindent=0.75em]
		\item \textbf{BRB-validity.~~} Suppose a correct process BRB-delivers message $m$ from a correct process $p_i$. Then, $p_i$ had BRB-broadcast $m$.
		\item \textbf{BRB-integrity.~~} No correct process BRB-delivers more than once.
		\item \textbf{BRB-no-duplicity.~~} No two correct processes BRB-deliver different messages from $p_i$ (who might be faulty).
		\item \textbf{BRB-Completion-1.~~} Suppose $p_i$ is a correct sender. All correct processes BRB-deliver from $p_i$ eventually.
		\item \textbf{BRB-Completion-2.~~} Suppose a correct process BRB-delivers a message from $p_i$ (who might be faulty). All correct processes BRB-deliver $p_i$'s message eventually.
	\end{itemize}
\end{definition}

\Subsubsection{\emsO{Repeated BRB.}} 
\emsO{Distributed systems use, over time, an unbounded number of BRB instances. We require our solution to use, at any given point in time, a bounded amount of memory. Thus, for} the sake of completeness, we also consider the problem of recycling an unbounded sequence of BRB invocations using bounded memory. We require the (single-instance) BRB object, $O$, to have an operation, called $\mathsf{recycle}()$, that allows the recycling mechanism locally reset $O$, after all non-faulty processes had completed the delivery of $O$'s message. Also, we require the mechanism to inform (the possibly recycled) $O$ regarding its availability to take new missions. Specifically, the $\mathsf{txAvailable}()$ operation returns $\true$ when the sender can use $O$ for broadcasting and $\mathsf{rxAvailable}()$ returns $\true$ when $O$'s new transmission has arrived at the receiver.

\emsO{One may observe that the problem statement does not depend on the fault model or the design criteria. However, the proposed solution depends on all three. To clarify, we solve the single instance BRB using the requirements presented by Raynal~\cite[Ch. 4]{DBLP:books/sp/Raynal18}. Then, we solve an extended version of the problem in which each BRB instance needs to be recycled so that an unbounded number of BRB instances can appear.}


\Subsection{Fault models} 
Recall that our BRB solution may be a component in a system that solves consensus. Thus, we safeguard against Byzantine failures by following the same assumptions that are often used when solving consensus. Specifically, for the sake of deterministic and signature-free solvability~\cite{DBLP:journals/jacm/PeaseSL80}, we assume there are at most $t<n/3$ crashed or Byzantine processes, where $n$ is the total number of processes. 
\emsO{The proposed solutions are for} message-passing systems that have no guarantees on the communication delay and without explicit access to the clock. \emsO{These systems are also prone to communication failures, \eg packet omission, duplication, and reordering, as long as fair communication (FC) holds, \ie if $p_i$ sends a message infinitely often to $p_j$, then $p_j$ receives that message infinitely often.
We use three different fault models with notations following Raynal~\cite{DBLP:books/sp/Raynal18}:} 

\begin{itemize}
	\item \emsO{$\mathsf{BAMP_{n,t}[FC, t < n/3]}$.~~ This is a Byzantine Asynchronous Message-Passing model with at most $t$ (out of $n$) faulty nodes. The array $[FC, t < n/3]$ denotes the list of all assumptions, \ie FC and $t<n/3$. We use this model for studying the problem of single-instance BRB since it has no synchrony assumptions.} 
	\item \emsO{$\mathsf{BAMP_{n,t}[FC,t < n/3, BML, \diamondsuit P_{mute}]}$.~~ By Doudou \etal~\cite{DBLP:conf/edcc/DoudouGGS99}, processes commit muteness failures when they stop sending specific messages, but they may continue to send ``I-am-alive'' messages. For studying the problem of BRB instance recycling, we enrich $\mathsf{BAMP_{n,t}[FC, t <}$ $\mathsf{n/3]}$ with a muteness detector of class $\diamondsuit P_{mute}$ and assume bounded message lifetime (BML). That is, in any unbounded sequence of BRB invocations, at the time that immediately follows the $x$-th invocation, the messages associated with the $(x\mathit{-}\lambda)$-th invocation (or earlier) are either delivered or lost, where $\lambda$ is a known upper-bound.}   
	\item \emsO{$\mathsf{AMP_{n}[FC, BML]}$.~~ 
	For the sake of a simple presentation of the repeated BRB solution, we first present a solution for the fault model, $\mathsf{AMP_{n}[FC, BML]}$, which does not consider any node failures (before presenting a repeated BRB solution for $\mathsf{BAMP_{n,t}[FC,t < n/3, BML, \diamondsuit P_{mute}]}$).}
\end{itemize}

\emsO{Raynal~\cite{DBLP:books/sp/Raynal18} refers to an asynchronous system as \textit{time-free} when it includes synchrony assumptions, \eg BML. 
Note that BML does not imply bounded communication delay since an unbounded number of messages can be lost between any two successful transmissions.
At last, our muteness detector implementation follows an assumption about the number, $\Theta$, of messages that some non-faulty processes can exchange without hearing from all non-faulty processes.}

\Subsection{Self-stabilization} 
%
%
Dijkstra's seminal work~\cite{DBLP:journals/cacm/Dijkstra74} demonstrated recovery within a finite time after the occurrence of the last transient fault, which may corrupt the system state in any manner (as long as the program code stays intact). Dijkstra offered an alternative to traditional fault-tolerance, which aims at assuring that the system, at all times, remains in a correct state under the assumption that the system state changes only due to the algorithmic steps and specified failures. Alas, the latter target is unattainable in the presence of failures that were unforeseen during the algorithm design. In order to address this concern, self-stabilization considers failures that are transient by nature and hard to be observed. Thus, they cannot be specified by the fault model, such as the one above, which includes process and communication failures. Therefore, self-stabilizing systems are required to recover eventually (in the presence of all foreseen and specified failures) after the occurrence of the last unforeseen and transient failure. 

In this paper, in addition to the faults specified above, we also aim to recover after the occurrence of the last \emph{arbitrary transient-fault}~\cite{DBLP:series/synthesis/2019Altisen,DBLP:books/mit/Dolev2000}. These transient-faults model any temporary violation of assumptions according to which the system was designed to operate. This includes the corruption of control variables, such as the program counter and packet payloads, as well as operational assumptions, such as that at most $t<n/3$ processes are faulty. Since the occurrence of these failures can be arbitrarily combined, we assume these transient-faults can alter the system state in unpredictable ways. When modeling the system, Dijkstra assumed that these violations can bring the system to an arbitrary state from which a \emph{self-stabilizing system} should recover~\cite{DBLP:journals/cacm/Dijkstra74}. \Ie Dijkstra requires the correctness proof of a self-stabilizing system to demonstrate recovery within a finite time after the last occurrence of a transient-fault and once the system has recovered, it must never violate the task requirements. \emsO{Arora and Gouda~\cite{DBLP:conf/ftcs/AroraG92} refer to the former property as convergence and the latter closure. Note that the stratification of the task requirements, which is Definition~\ref{def:prbDef}  in the case of this paper, holds only when closure is guaranteed. To say it in other words, only after the system has finished recovering from the occurrence of the last transient fault does a self-stabilizing system guarantees the satisfaction of the task requirement, for details see~\cite{DBLP:series/synthesis/2019Altisen,DBLP:books/mit/Dolev2000}.}

\Subsection{Related work}
In the context of reliable broadcast, there are (non-self-stabilizing) Byzantine fault-tolerant (BFT) solutions~\cite{DBLP:books/sp/Raynal18,DBLP:conf/icdcs/BonomiDFRT21,DBLP:conf/opodis/GuerraouiKKPST20,DBLP:conf/sss/AlbouyFRT21} and (non-BFT) self-stabilizing solutions~\cite{DBLP:conf/netys/LundstromRS20} (even for total order broadcast~\cite{DBLP:conf/netys/LundstromRS20,DBLP:conf/icdcn/LundstromRS21,lundstrom2021self,TR}).
We focus on BT~\cite{DBLP:journals/jacm/BrachaT85,DBLP:conf/podc/BrachaT83} to which we propose a self-stabilizing variation. BT is the basis for advanced BFT algorithms for solving consensus~\cite{DBLP:journals/tpds/MostefaouiR16} and is based on a simpler communication abstraction by Toueg that is called no-duplicity broadcast~\cite{DBLP:conf/podc/Toueg84}. It includes all of Definition~\ref{def:prbDef}'s requirements except for BRB-Completion-2.
Maurer and Tixeuil~\cite{DBLP:conf/srds/MaurerT14} consider an abstract that is perhaps simpler than no-duplicity since they only consider no-duplicity (and none of the other requirements of Definition~\ref{def:prbDef}). They provide a single-instance synchronous self-stabilizing BFT broadcast, whereas we consider an asynchronous repeated BRB that follows Definition~\ref{def:prbDef}, \emsO{which is taken from Raynal~\cite{DBLP:books/sp/Raynal18}. Raynal studies the exact power of all the essential communication abstractions in the area of fault-tolerant message-passing systems. We study the more useful definition provided by Raynal since we wish to connect our solution to all relevant protocols in the area.}
	

Our study focuses on the BT~\cite{DBLP:journals/jacm/BrachaT85,DBLP:conf/podc/BrachaT83} solution to which we propose a self-stabilizing variation. BT is the basis for advanced BFT algorithms for solving consensus, such as the one by Most{\'{e}}faoui and Raynal~\cite{DBLP:journals/tpds/MostefaouiR16}. BT is based on a simpler communication abstraction called no-duplicity broadcast (ND-broadcast) by Toueg~\cite{DBLP:conf/podc/Toueg84,DBLP:books/sp/Raynal18}. It includes all of the above requirements except BRB-Completion-2.

In the broader context of self-stabilizing BFT solutions for message-passing systems, we find solutions for topology discovery~\cite{DBLP:conf/netys/DolevLS13}, storage~\cite{DBLP:conf/sss/BonomiPPT18,DBLP:conf/srds/BonomiPPT17,DBLP:conf/podc/BonomiPPT16,DBLP:conf/ipps/BonomiPT15,DBLP:conf/podc/BonomiDPR15}, clock synchronization~\cite{DBLP:conf/podc/DolevW95,DBLP:journals/jacm/LenzenR19,DBLP:journals/mst/KhanchandaniL19}, approximate agreement \cite{DBLP:journals/tcs/BonomiPPT19}, asynchronous unison~\cite{DBLP:journals/jpdc/DuboisPNT12}, communication in dynamic networks~\cite{DBLP:conf/opodis/Maurer20} to name a few. Also, BFT state-machine replication by Binun \etal~\cite{DBLP:conf/sss/BinunCDKLPYY16,DBLP:conf/cscml/BinunDH19} for synchronous systems and Dolev \etal~\cite{DBLP:conf/cscml/DolevGMS18} for practically-self-stabilizing partially-synchronous systems.

Even though Doudou \etal consider the consensus problem while we consider here repeated BRB, both works share the same motivation, \ie circumventing known impossibilities, \eg the one by Fischer, Lynch, and Paterson~\cite{DBLP:journals/jacm/FischerLP85}.




\Subsection{Our contribution}
We present a fundamental module for dependable distributed systems: SSBRB, a self-stabilizing BFT reliable broadcast for asynchronous message-passing systems, \ie for the model of $\mathsf{BAMP_{n,t}[FC,t < n/3]}$. We obtain this new self-stabilizing solution via a transformation of the non-self-stabilizing BT algorithm~\cite{DBLP:journals/jacm/BrachaT85,DBLP:conf/podc/BrachaT83} while preserving BT's resilience optimality of $t < n/3$.

In the absence of transient-faults, our asynchronous solution for single-instance BRB achieves operation completion within a constant number of communication rounds. After the occurrence of the last transient-fault, the system recovers eventually (while assuming execution fairness among the non-faulty processes). The amount of memory used by the proposed algorithm is bounded and the communication costs of the studied and proposed algorithms are similar, \ie $\bigO(n^2)$ messages per BRB instance. The main difference is that our solution unifies all the types of messages sent by BT into a single message that is repeatedly sent. This repetition is imperative since self-stabilizing systems cannot stop sending messages~\cite[Chapter 2.3]{DBLP:books/mit/Dolev2000}. 

Our contribution also includes a self-stabilizing BFT recycling mechanism for time-free systems that are enriched with muteness detectors, \ie $\mathsf{BAMP_{n,t}[FC,}$ $\mathsf{ BML, \diamondsuit P_{mute}]}$. The mechanism is based on an algorithm that counts communication rounds. Since individual BRB-broadcasters increment the counter independently, the algorithm is named the \emph{independent round counter} (IRC) algorithm. Implementing a self-stabilizing BFT IRC is a non-trivial challenge since this counter should facilitate an unbounded number of increments, yet it has to use only a constant amount of memory. Using novel techniques for dealing with integer overflow events, the proposed solution recovers from transient faults eventually, uses a bounded amount of memory, and has communication costs of $\bigO(n)$ messages per BRB instance.

To the best of our knowledge, we propose the first self-stabilizing BFT solutions for the problems of IRC and repeated BRB (that follows BT's problem specifications~\cite[Ch. 4]{DBLP:books/sp/Raynal18}). As said, BRB and IRC consider different fault models. Section~\ref{sec:sys} defines $\mathsf{BAMP_{n,t}[FC, t < n/3]}$ and self-stabilization. The non-self-stabilizing BT algorithm for $\mathsf{BAMP_{n,t}[FC, t < n/3]}$ is studied in Section~\ref{sec:BRB}. Our self-stabilization BFT variation on BT for $\mathsf{BAMP_{n,t}[FC,}$  $\mathsf{t < n/3]}$ is proposed in Section~\ref{sec:singleBRB}. IRC is presented in two steps. A self-stabilizing IRC for time-free node-failure-free message-passing systems appears in Section~\ref{sec:sysFree}. In Section~\ref{sec:sysFreeSystem}, we revise these time-free settings into $\mathsf{BAMP_{n,t}[FC,t < n/3, BML, \diamondsuit P_{mute}]}$ and propose a self-stabilizing BFT IRC. Section~\ref{sec:con} compares the overhead of the studied and proposed solutions when executing $\delta$ BRB instances concurrently. This straightforward extension is imperative for the sake of practical deployments. 

Our \emsO{specifications (Definition~\ref{def:prbDef}) follow the ones by Raynal~\cite[Ch. 4]{DBLP:books/sp/Raynal18}. Thus, our} SSBRB solution can serve as a building block for multivalued consensus~\cite{DBLP:journals/corr/abs-2110-08592}.

\medskip
For the reader's convenience, we include a Glossary just before the section of References.

\Section{System Settings for $\mathsf{BAMP_{n,t}[FC,t < n/3]}$}
%
\label{sec:sys} 
\emsO{This work focuses on three fault models, \ie $\mathsf{BAMP_{n,t}[FC,t < n/3]}$, which we present in this section, as well as the fault model of $\mathsf{AMP_{n}[FC, BML]}$ and the fault model of $\mathsf{BAMP_{n,t}[FC,t < n/3, BML, \diamondsuit P_{mute}]}$, which we present in sections~\ref{sec:SelfIRC} and~\ref{sec:sysFreeSystem}, respectively. The model considered in this section is for} asynchronous message-passing systems that have no guarantees on the communication delay. Also, the algorithm cannot explicitly access the (local) clock (or use timeout mechanisms). The system consists of a set, $\sP$, of $n$ fail-prone nodes (or processes) with unique identifiers. Any pair of nodes $p_i,p_j \in \sP$ has access to a bidirectional communication channel, $\mathit{channel}_{j,i}$, that, at any time, has at most $\capacity \in \bZ^+$ messages on transit from $p_j$ to $p_i$ (this assumption is due to a known impossibility~\cite[Chapter 3.2]{DBLP:books/mit/Dolev2000}).

%
\label{sec:interModel}
In the \emph{interleaving model}~\cite{DBLP:books/mit/Dolev2000}, the node's program is a sequence of \emph{(atomic) steps}. Each step starts with an internal computation and finishes with a single communication operation, \ie a message $send$ or $receive$. 
The \emph{state}, $s_i$, of node $p_i \in \sP$ includes all of $p_i$'s variables and \emsO{all incoming communication channels, $\mathit{channel}_{j,i}:p_i,p_j\in \sP$}. The term \emph{system state} (or configuration) refers to the tuple $c = (s_1, s_2, \cdots,  s_n)$. We define an \emph{execution (or run)} $R={c[0],a[0],c[1],a[1],\ldots}$ as an alternating sequence of system states $c[x]$ and steps $a[x]$, such that each $c[x+1]$, except for the starting one, $c[0]$, is obtained from $c[x]$ by $a[x]$'s execution.


\Subsection{The fault model and self-stabilization}
The \emph{legal executions} ($LE$) set refers to all the executions in which the requirements of task $T$ hold. In this work, $T_{\text{BRB}}$ denotes the task of BFT Reliable Broadcast, which Section~\ref{sec:intro} specifies, and the executions in the set $LE_{\text{BRB}}$ fulfill $T_{\text{BRB}}$'s requirements.

\Subsubsection{Benign failures}
\label{sec:benignFailures}
A failure occurrence is a step that the environment takes rather than the algorithm. When the occurrence of a failure cannot cause the system execution to lose legality, \ie to leave $LE$, we refer to that failure as a benign one. 

\BSubsubsubsection{Communication failures and fairness.}
We focus on solutions that are oriented towards asynchronous message-passing systems and thus they are oblivious to the time at which the packets depart and arrive. We assume that any message can reside in a communication channel only for a finite period. Also, the communication channels are prone to packet failures, such as loss, duplication, and reordering.  However, if $p_i$ sends a message infinitely often to $p_j$, node $p_j$ receives that message infinitely often. We refer to the latter as the \emph{fair communication} assumption. 
%
%
%
As in~\cite{DBLP:conf/netys/GeorgiouMRS21}, we assume that the communication channel from a correct node eventually includes only messages that were transmitted by the sender.   

The studied algorithm assumes reliable communication channels whereas the proposed solution does not make any assumption regarding reliable communications. Section~\ref{sec:intermediateMMR} provides further details regarding the reasons why the proposed solution cannot make this assumption.

\BSubsubsubsection{Arbitrary node failures.}
\label{sec:arbitraryNodeFaults}
Byzantine faults model any fault in a node including crashes, and arbitrary malicious behaviors. Here the adversary lets each node receive the arriving messages and calculate their state according to the algorithm. However, once a node (that is captured by the adversary) sends a message, the adversary can modify the message in any way, delay it for an arbitrarily long period or even omit it from the communication channel. The adversary can also send fake messages, \ie not according to the algorithm. Note that the adversary has the power to coordinate such actions without any computational (or communication) limitation. 
%
%
For the sake of solvability~\cite{DBLP:journals/toplas/LamportSP82,DBLP:journals/jacm/PeaseSL80,DBLP:conf/podc/Toueg84}, the fault model that we consider limits only the number of nodes that can be captured by the adversary. That is, the number, $t$, of Byzantine failures needs to be less than one-third of the number, $n$, of nodes in the system, \ie $3t+1\leq n$. The set of non-faulty indexes is denoted by $\Correct$, so that $i \in \Correct$ when $p_i$ is a correct node.

\Subsubsection{Arbitrary transient-faults}
\label{sec:arbitraryTransientFaults}
We consider any temporary violation of the assumptions according to which the system was designed to operate. We refer to these violations and deviations as \emph{arbitrary transient-faults} and assume that they can corrupt the system state arbitrarily (while keeping the program code intact). The occurrence of a transient fault is rare. Thus, we assume that the last arbitrary transient fault occurs before the system execution starts~\cite{DBLP:books/mit/Dolev2000}. Also, it leaves the system to start in an arbitrary state.

\Subsection{Dijkstra's self-stabilization}
\label{sec:Dijkstra}
An algorithm is \emph{self-stabilizing} with respect to $LE$, when every execution $R$ of the algorithm reaches within a finite period a suffix $R_{legal} \in LE$ that is legal. Namely, Dijkstra~\cite{DBLP:journals/cacm/Dijkstra74} requires $\forall R:\exists R': R=R' \circ R_{legal} \land R_{legal} \in LE \land |R'| \in \bZ^+$, where the operator $\circ$ denotes that $R=R' \circ R''$ is the concatenation of $R'$ with $R''$. 


	\Subsection{Wait-free guarantees and transient-faults recovery assuming seldom fairness}
\label{sec:fairnessEx}
Wait-free algorithms guarantee that operations (that were invoked by non-failing nodes) always complete in the presence of asynchrony and $t$ faulty nodes. Self-stabilizing algorithms sometimes assume that their executions are fair~\cite{DBLP:books/mit/Dolev2000}. That is, given a step $a$, we say that $a$ is \emph{applicable} to system state $c$ if there exists system state $c'$, such that $a$ leads to $c'$ from $c$. We say that a system execution is \emph{fair} when every step of a correct node that is applicable infinitely often is executed infinitely often and fair communication is kept. This work assumes execution fairness during the period in which the system recovers from the occurrence of the last arbitrary transient fault. Since the occurrence of transient faults is rare, only seldom do our fairness assumptions needed and just for the period of recovery. The rest of the time, \ie in the absence of transient faults or after the recovery from them, the execution is assumed to be arbitrary.


%

\Section{The non-self-stabilizing BT algorithm}
%
\label{sec:BRB}
%
%
%
Recall that the studied algorithm, BT~\cite{DBLP:journals/jacm/BrachaT85,DBLP:conf/podc/BrachaT83}, is a BRB solution for $\mathsf{BAMP_{n,t}[FC,t <}$ $\mathsf{n/3]}$. BT is based on a simpler communication abstraction called no-duplicity broadcast (ND-broadcast) by Toueg~\cite{DBLP:conf/podc/Toueg84,DBLP:books/sp/Raynal18}. The ND-broadcast task includes all of the BRB requirements (Section~\ref{sec:problem}) except BRB-Completion-2. We review BT after studying Toueg's ND-broadcast algorithm.

\begin{algorithm}[t!]
	\begin{\algSize}	
		
		\smallskip
		
		\textbf{operation} $\mathsf{ndBroadcast}(m)$
		\label{ln:NDbrbBroadcast}\textbf{do} \textbf{broadcast}\label{ln:NDndINITsend} $\mathrm{INIT}(m)$\;
		
		\smallskip
		
		\textbf{upon} $\mathrm{INIT}(\mathit{mJ})$ \textbf{first arrival from} $p_j$ \textbf{do} \textbf{broadcast}\label{ln:NDndECHOsend} $\mathrm{ECHO}(j,\mathit{mJ})$\;
		
		\smallskip
		
		\textbf{upon} $\mathrm{ECHO}(\mathit{k},\mathit{mJ})$ \textbf{arrival from}\label{ln:NDuponEcho} $p_j$ \Begin{\If{$\mathrm{ECHO}(\mathit{k},\mathit{mJ})$ received from at least $(n\mathit{+}t)/2$ nodes}{$\mathrm{ndDeliver}(j,\mathit{mJ})$;\label{ln:NDechoThen} } }

		\smallskip
		
		\caption{\label{alg:nd}non-self-stabilizing no-duplicity broadcast (ND-broadcast); code for $p_i$.}
	\end{\algSize}
\end{algorithm}

\begin{algorithm}[t!]
	\begin{\algSize}	
		
		\smallskip
		
		\textbf{operation} $\mathsf{brbBroadcast}(m)$
		\label{ln:brbBroadcast}\textbf{do} \textbf{broadcast}\label{ln:ndINITsend} $\mathrm{INIT}(m)$\;
		
		\smallskip
		
		\textbf{upon} $\mathrm{INIT}(\mathit{mJ})$ \textbf{first arrival from} $p_j$ \textbf{do} \textbf{broadcast}\label{ln:ndECHOsend} $\mathrm{ECHO}(j,\mathit{mJ})$\;
		
		\smallskip
		
		\textbf{upon} $\mathrm{ECHO}(\mathit{k},\mathit{mJ})$ \textbf{arrival from}\label{ln:uponEcho} $p_j$ \Begin{ \lIf{$\mathrm{ECHO}(\mathit{k},\mathit{mJ})$ received from at least $(n\mathit{+}t)/2$ nodes 
				$\land  \; \mathrm{READY}(\mathit{k},\mathit{mJ})$ not yet broadcast}{\label{ln:echoThen} \textbf{broadcast} $\mathrm{READY}(\mathit{k},\mathit{mJ})$ } }
		
		\smallskip
		
		\textbf{upon} $\mathrm{READY}(\mathit{k},\mathit{mJ})$ \textbf{arrival from} $p_j$\label{ln:uponReady} \Begin{
			
			\lIf{$\mathrm{READY}(\mathit{k},\mathit{mJ})$ received from $(t\mathit{+}1)$ nodes $\land$ $\mathrm{READY}(\mathit{k},\mathit{mJ})$ not yet broadcast\label{ln:t1Then}}{\textbf{broadcast} $\mathrm{READY}(\mathit{k},\mathit{mJ})$\label{ln:t1Else}}
			
			\lIf{$\mathrm{READY}(\mathit{k},\mathit{mJ})$ received from at least $(2t\mathit{+}1)$ nodes $\land \langle \mathit{k},\mathit{mJ} \rangle$ not yet BRB-Delivered 
					\label{ln:21t1Then}}{\textbf{brbDeliver} $(\mathit{k},\mathit{mJ})$\label{ln:ndbrbDeliver}}
			
		}		
		
		\smallskip
		
		\caption{\label{alg:brb}non-self-stabilizing Byzantine Reliable Broadcast (BRB); code for $p_i$.}
	\end{\algSize}
\end{algorithm}


\Subsection{No-Duplicity Broadcast}
Algorithm~\ref{alg:nd} brings Toueg's solution for ND-broadcast~\cite{DBLP:conf/podc/Toueg84}. 
%
%
Algorithm~\ref{alg:brb} assumes that every correct node invokes ND-broadcast at most once. Node $p_i$ initiates the ND-broadcasts of $m_i$ by sending $\mathrm{INIT}(m_i)$ to all nodes (line~\ref{ln:NDndINITsend}). Upon this message's first arrival to node $p_j$, it disseminates the fact that $p_i$ has initiated $m$'s ND-broadcast by sending $\mathrm{ECHO}(i,m)$ to all nodes (line~\ref{ln:NDndECHOsend}). Upon this message arrival to $p_k$ from more than $(n\mathit{+}t)/2$ different nodes, $p_k$ is ready to ND-deliver $\langle i,m_i \rangle$ (line~\ref{ln:NDuponEcho}).

\Subsection{Byzantine Reliable Broadcast}
As explained, we present the BT solution for BRB as an extension of Toueg's solution for ND-broadcast. Algorithm~\ref{alg:brb} satisfies the BRB requirements (Section~\ref{sec:problem}) assuming $t < n/3$. Note that the line numbers of Algorithm~\ref{alg:brb} continue the ones of Algorithm~\ref{alg:nd}. 


The first difference between the ND and BRB algorithms is in the consequent clause of the if-statement in line~\ref{ln:echoThen}, where ND-delivery of $\langle j,m\rangle$ is replaced with the broadcast of $\mathrm{READY}(j,m)$. This broadcast indicates that $p_i$ is ready to BRB-deliver $\langle j,m\rangle$ as soon as it receives sufficient support, \ie the arrival of $\mathrm{READY}(j,m)$, which tells that $\langle j,m\rangle$ can be BRB-delivered. Note that BRB-no-duplicity protects Algorithm~\ref{alg:brb} from the case in which $p_i$ broadcasts $\mathrm{READY}(j,m)$ while $p_j$ broadcasts $\mathrm{READY}(j,m')$, such that $m \neq m'$.

The new part of the BRB algorithm (lines~\ref{ln:uponReady} to~\ref{ln:ndbrbDeliver}) includes two if-statements. The first one (line~\ref{ln:t1Then}) makes sure that every correct node receives $\mathrm{READY}(j,m)$ from at least one correct node before BRB-delivering $\langle j,m\rangle$. This is done via the broadcasting of $\mathrm{READY}(j,m)$ as soon as $p_i$ received it from at least $(t \mathit{+} 1)$ different nodes (since $t$ of them can be Byzantine).

The second if-statement (line~\ref{ln:ndbrbDeliver}) makes sure that no two correct nodes BRB-deliver different pairs (in the presence of plausibly fake $\mathrm{READY}(j,\bull)$ messages sent by Byzantine nodes, where the symbol `$\bull$' stands for any legal value). That is, the delivery of a BRB-broadcast is done only after the first reception of the pair $\langle j,m\rangle$ from at least $(2t\mathit{+}1)$ (out of which at most $t$ are Byzantine). The receiver then knows that there are at least $t\mathit{+}1$ correct nodes that can make sure that the condition in line~\ref{ln:t1Then} holds eventually for all correct nodes.


\Section{Self-stabilizing Byzantine-tolerant Single-instance BRB}
\label{sec:singleBRB}
Before proposing our solution (Section~\ref{sec:propBRB}), we review the challenges that we face when transforming the non-self-stabilizing BT algorithm~\cite{DBLP:journals/jacm/BrachaT85,DBLP:conf/podc/BrachaT83} into a self-stabilizing one (Section~\ref{sec:challenge}).

\Subsection{Challenges and approaches}
\label{sec:challenge}
We analyze the behavior of the BT algorithm in the presence of transient-faults. We clarify that our analysis is relevant only in the context of self-stabilization since Bracha and Toueg do not consider transient-faults.

\Subsubsection{Dealing with memory corruption and desynchronized system states}
Recall that transient faults can corrupt the system state in any manner (as long as the program code remains intact). For example, memory corruption can cause the local state to indicate that a certain message has already arrived (line~\ref{ln:ndECHOsend}) or that a certain broadcast was already performed (line~\ref{ln:t1Then}). This means that some necessary messages will not be broadcast. This will result in an indefinite blocking. The proposed solution avoids such a situation by unifying all messages into a single $\mathrm{MSG}(\mathit{mJ})$, where the field $\mathit{mJ}$ includes all the fields of the messages of Algorithm~\ref{alg:brb}.

\Subsubsection{Datagram-based end-to-end communications}
\label{sec:intermediateMMR}
Algorithm~\ref{alg:brb} assumes reliable communication channels when broadcasting in a quorum-based manner, \ie sending the same message to all nodes and then waiting for a reply from $n\mathit{-}f$ nodes. Next, we explain why, for the sake of a simpler presentation, we choose not to follow this assumption.
Self-stabilizing end-to-end communications require a known bound on the capacity of the communication channels~\cite[Chapter 3]{DBLP:books/mit/Dolev2000}. In the context of self-stabilization and quorum systems, we must avoid situations in which communicating in a quorum-based manner can lead to a contradiction with the system assumptions. Dolev, Petig, and Schiller~\cite{DBLP:journals/corr/abs-1806-03498} explain that there might be a subset of nodes that are able to complete many round-trips with a given sender, while other nodes merely accumulate messages in their communication channels. The channel bounded capacity implies that the system has to either block or omit messages before their delivery. Thus, the proposed solution does not assume access to reliable channels. Instead, communications are simply repeated by the algorithm's do-forever loop.



\begin{algorithm}[t!]
	\begin{\algSize}	
		
		\smallskip
		
		\textbf{types:}
		\label{ln:types} $\texttt{brbMSG} :=\{\init, \echo, \ready\}$; 
		%
		
		
		\smallskip
		
		\textbf{variables:}
		%
		%
		$msg[\sP][\texttt{brbMSG}]:=[[\emptyset,\ldots,\emptyset]]$ \texttt{/*} most recently sent/received message \texttt{*/}
		
		\emsO{$wasDelivered[\sP]:=[\false,\ldots,\false]$ \texttt{/*} indicates whether the message was delivered \texttt{*/}}

		
		\smallskip
		
		%
		%
		%
		%
		%
		
		\textbf{required interfaces:}  	
		$\mathsf{txAvailable}()$ and $\mathsf{rxAvailable}(k)$\label{ln:start}
		%
		
		
		\smallskip
		
		\textbf{provided interfaces:}  	
		$\mathsf{recycle}(k)$ \label{ln:reset}\textbf{do} \emsO{$\{(msg[k],wasDelivered[k]) \gets ([\emptyset,\emptyset,\emptyset],\false)\}$\;}

		\smallskip
		
		$\mathsf{mrg}(\mathit{mJ},j)$ \label{ln:merge}\Begin{
			\ForEach{$s \in \emph{\texttt{brbMSG}}, p_k \in \sP$}{
				\lIf{$s \neq \emph{\init} \lor $ $\nexists s = \emph{\init},(k,m),(k,m') \in$ $(msg[j][s] \cup \mathit{mJ}[s]): m$$\neq$$m'$\label{ln:messageConsistentIf}}{$msg[j][s]$ $\gets$ $msg[j][s] \cup$ $\mathit{mJ}[s]$\label{ln:messageConsistentThen}}
			}
		}
		
		\smallskip
		
		\textbf{operations:}  		
		
		$\mathsf{brbBroadcast}(v)$ \label{ln:brbBradcast} \textbf{do} \{\textbf{if} {$\mathsf{txAvailable}()$} \textbf{then} {$\mathsf{recycle}(i)$; $msg[i][\init]\gets\{v\}$}\}		
		
		\smallskip
		
		
		$\mathsf{brbDeliver}(k)$ \label{ln:brbDeliver}\Begin{
			\If{$\exists m: (2t\mathit{+}1) \leq |\{ p_{\ell} \in \sP: {(k,m) \in msg[\ell][\emph{\ready}]}\}|\land \mathsf{rxAvailable}(k)$}{\emsO{$wasDelivered[k]\gets wasDelivered[k] \land m\neq \bot$;} \Return{$m$}\label{ln:brbDeliverM}}\lElse{\Return{$\bot$}\label{ln:brbDeliverBot}}
		}
		
		\smallskip
		
		\emsO{$\mathsf{brbWasDelivered}(k)$ \label{ln:brbWasDelivered}\textbf{do} \Return{$wasDelivered[k]$}\;}

		\smallskip
		
		\textbf{do-forever} \Begin{
			
			{
				
				\If{$\exists {(j,m) \in msg[i][\emph{\echo}]}: m \notin msg[j][\emph{\init}] \lor \exists {(j,m) \in msg[i][\emph{\ready}]}: \neg ((n\mathit{+}t)/2$ $< |\{ p_{\ell} \in  \sP : (j,m) \in msg[\ell][\emph{\echo}]\}| \lor (t\mathit{+}1) \leq |\{ p_{\ell} \in  \sP: (j,m) \in msg[\ell][\emph{\ready}]\}|)$}{$\mathsf{recycle}(i)$\label{ln:consistent1}}
				
				\ForEach{$p_k \in \sP$}{
					
					\lIf{$|msg[k][\emph{\init}]| > 1 \lor \exists {s \neq \init} : \exists {p_j \in \sP} :$ $\exists {(j,m),(j,m') \in msg[k][s]}:$ $m\neq m'$}{$msg[k][s] \gets \emptyset$\label{ln:consistent0}}				
					
					\lIf{$\exists { m \in msg[k][\emph{\init}]}:msg[i][\emph{\echo}]=\emptyset$}{$msg[i][\echo] \gets \{(k,m)\}$\label{ln:initEcho}}
					
					\If{$\exists { m} :(n\mathit{+}t)/2<|\{ p_{\ell} \in  \sP: (k,m) \in msg[\ell][\emph{\echo}]\}|$\label{ln:echoReady0If}}{$msg[i][\ready] \gets msg[i][\ready] \cup \{(k,m)\}$\label{ln:echoReady0}}
					
					\If{$\exists m: (t\mathit{+}1) \leq |\{ p_{\ell} \in  \sP: (k,m) \in msg[\ell][\emph{\ready}]\}|$}{$msg[i][\ready] \gets msg[i][\ready] \cup \{(k,m)\}$\label{ln:echoReady1}}
					
				}	
			}

			
			\textbf{broadcast} $\mathrm{MSG}(\mathit{brb}I=msg[i],$ \fbox{$\mathit{irc}I=\mathsf{txMSG}())$;\label{ln:broadcast}}
			
		}

		

		
		\textbf{upon} $\mathrm{MSG}(\mathit{brb}J,$ \fbox{$\mathit{irc}J)$\label{ln:BRBupon}} \textbf{arrival from} $p_j$  \Begin{$\mathit{mrg}(\mathit{brb}J,j)$\;\fbox{$\mathsf{rxMSG}(\mathit{brb}J,\mathit{irc}J,j)$}}
		
		
		\medskip
		
		\caption{\label{alg:consensus}Self-stabilizing BFT BRB with instance recycling interface; $p_i$'s code}
	\end{\algSize}
	
\end{algorithm}

\Subsection{Self-stabilizing BFT single-instance solution}
\label{sec:propBRB}
%
%
Algorithm~\ref{alg:consensus} proposes our SSBRB solution for $\mathsf{BAMP_{n,t}[FC,t < n/3]}$. \emsO{The key idea is to (i) offer a variance of Algorithm~\ref{alg:brb} that its operations always complete even when starting from a corrected state, (ii) offer interfaces for coordinating the recycling of a given BRB object, as well as (iii) offer interfaces for accessing the delivered value and current status of the broadcast. This way, the recycling coordination mechanism (Section~\ref{sec:sysFree} and Section~\ref{sec:sysFreeSystem}) can make sure that no BRB object is recycled before all correct nodes deliver its result. Also, once all correct nodes have delivered a message, the BRB object can be recycled eventually.} The line numbers of Algorithm~\ref{alg:consensus} continue the ones of Algorithm~\ref{alg:brb}. The \fbox{boxed} code fragments in lines~\ref{ln:broadcast} to~\ref{ln:BRBupon} are irrelevant to our single-instance BRB implementation.


\Subsubsection{Types, constants, variables, and message structure.~~}
As mentioned, the message $\mathrm{MSG}()$ unifies the messages of Algorithm~\ref{alg:brb}. The array $msg[][]$ stores both the information that is sent and arrived by these messages. Specifically, $msg_i[i][]$ stores the information that node $p_i$ broadcasts (line~\ref{ln:broadcast}) and for any $j\neq i$ the entry $msg_i[j][]$ stores the information coming from $p_j$ (lines~\ref{ln:messageConsistentIf} to~\ref{ln:messageConsistentThen}). Also, we define the type $\texttt{brbMSG} :=\{\init, \echo, \ready\}$ (line~\ref{ln:types}) for storing information related to BRB-broadcast messages, \eg $msg_i[i][\init]$ stores the information that BRB-broadcast disseminates of $\mathrm{INIT}()$ messages and the results of the content of $\mathrm{READY}()$ messages appear in $msg_i[i][\ready]$. 


\Subsubsection{Algorithm details.~~}
The $\mathsf{brbBroadcast}(v)$ operation (line~\ref{ln:brbBradcast}) allows Algorithm~\ref{alg:consensus} to invoke BRB-broadcast instances with $v$. Such an invocation causes Algorithm~\ref{alg:consensus} to follow the logic of the BRB solution presented by Algorithm~\ref{alg:brb} in lines~\ref{ln:brbDeliver} and~\ref{ln:initEcho} to~\ref{ln:echoReady1}. We note that our solution also includes consistency tests at line~\ref{ln:consistent1}.

\begin{figure}
	\begin{center}
		\includegraphics[scale=0.5, clip]{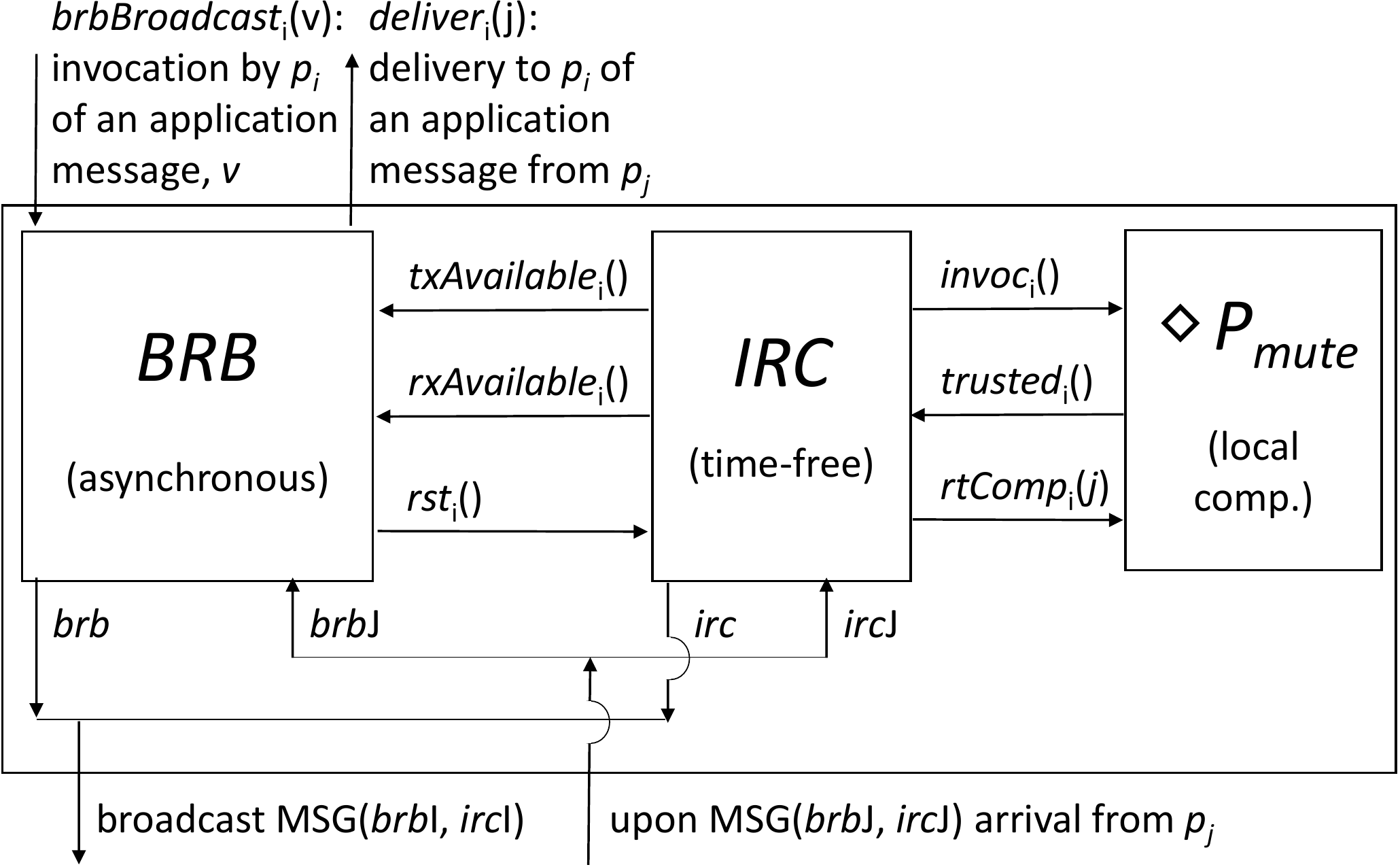}
	\end{center}
	\caption{\label{fig:BrbIRC}\small{Integrating the BRB (Section~\ref{sec:singleBRB}), IRC (Section~\ref{sec:sysFree}), and $\diamondsuit P_{mute}$  (Section~\ref{sec:sysFreeSystem}) protocols}}
\end{figure}

\Subsubsection{Interfaces \emsO{for coordinating the recycling of a given BRB object.~~}}
%
%
Recall that Algorithm~\ref{alg:consensus} has an interface to a recycling mechanism of BRB instances (Section~\ref{sec:sysFree}). The interface between the proposed BRB and recycling mechanism includes the $\mathsf{recycle}()$, $\mathsf{txAvailable}()$, and $\mathsf{rxAvailable}()$ operations, see Figure~\ref{fig:BrbIRC} (the interface between IRC and $\diamondsuit P_{mute}$ is irrelevant to Algorithm~\ref{alg:consensus}). The function $\mathsf{recycle}_i(k)$ (line~\ref{ln:reset}) lets the recycling mechanism locally reset $msg_i[k][]$ with the notation $f_i()$ denoting that $p_i$ executes the function $f()$. For the single-instance BRB (without recycling), define $\mathsf{txAvailable}()$ and $\mathsf{rxAvailable}(k)$ (line~\ref{ln:start}) to return $\true$. Note that we further integrate between BRB and IRC is via the piggybacking of their messages.


\Subsubsection{Interfaces \emsO{for accessing the delivered value and current status.~~}}
\emsO{Algorithms~\ref{alg:nd} and~\ref{alg:brb} inform the application layer about message arrival by raising the events of $\mathrm{ndDeliver}()$ (line~\ref{ln:NDechoThen}), and respectively, $\mathsf{brbDeliver}()$ (line~\ref{ln:ndbrbDeliver}). Our SSBFT BRB solution takes another approach in which the application is pulling information from Algorithm~\ref{alg:consensus} by invoking the $\mathsf{brbDeliver}()$ operation (line~\ref{ln:brbDeliver}), which returns $\bot$ (line~\ref{ln:brbDeliverBot}) when no message is ready to be delivered. Otherwise, the arriving message is returned (line~\ref{ln:brbDeliverM}). Note that once $\mathsf{brbDeliver}_i(k):i,k\in \Correct$ returns a non-$\bot$ value, $\mathsf{brbDeliver}_i(k)$ returns a non-$\bot$ value in all subsequent invocations. For the sake of satisfying BRB-integrity (Definition~\ref{def:prbDef}) in a self-stabilizing manner, line~\ref{ln:brbDeliverM} records the fact that the non-$\bot$ message was delivered at least once by storing $\true$ in $wasDelivered_i[k]$. The application can access the value stored in $wasDelivered_i[k]$ by invoicing $\mathsf{brbWasDelivered}_i(k)$ (line~\ref{ln:brbWasDelivered}).}

\Section{Correctness of Algorithm~\ref{alg:consensus}}
%
Definition~\ref{def:consistent} defines the terms active nodes and consistent executions. 
Theorem~\ref{thm:recoveryconsensusI} shows that consistency is regained eventually. 
Then, we provide a proof of completion (Theorem~\ref{thm:brbTerminateSimple}) before demonstrating the closure properties (Theorem~\ref{thm:brbClousre}). The closure proof (Section~\ref{sec:brbClosure}) shows that the proposed solution satisfies BRB task requirements (Definition~\ref{def:prbDef}). It is based on the assumption that BRB objects are eventually recycled after their task was completed (Section~\ref{sec:problem}).  


\emsO{Definition~\ref{def:consistent} considers the if-statement conditions of lines~\ref{ln:consistent1} and~\ref{ln:consistent0}, see items (brb.i) and (brb.ii). Item (brb.iii) has similar considerations as the ones of Item (brb.i) in the context of the sent messages.}


\begin{definition}[Active nodes and consistent executions of Algorithm~\ref{alg:consensus}]
	\label{def:consistent}
	%
	%
	We use the term \emph{active} for node $p_i \in \sP$ when referring to the case of $msg_i[i][\emph{\init}]$ $\neq \emptyset$.
	Let $R$ be an Algorithm~\ref{alg:consensus}'s execution, $p_i, p_j\in \sP: i \in \Correct$, and $c \in R$. Suppose in $c$: 
	
	\begin{itemize}
		\item \textbf{(brb.i)} $|msg_i[j][\emph{\init}]|\leq 1$ and $\nexists_{t \in \{\emph{\echo}, \emph{\ready}\}}$ $\exists_{p_k \in \sP}$ $\exists_{(k,m),(k,m') \in msg_i[j][t]}$ $m\neq m'$.
		
		\item \textbf{(brb.ii)} $\forall_{(k,m) \in msg_i[i][\emph{\ready}]}  ((n\mathit{+}t)/2< |\{ p_{\ell} \in  \sP : (k,m) \in msg_i[\ell][\emph{\echo}]\}| \lor (t\mathit{+}1) \leq |\{ p_{\ell} \in  \sP: (k,m) \in msg_i[\ell][\emph{\ready}]\}|)$.				
		
		\item \textbf{(brb.iii)} for any message $\mathsf{MSG}(\mathit{brb}J=\mathit{mJ},\bull)$ in transient from $p_i$ to $p_j$, it holds that for any ${p_k \in \sP}$ and ${t \neq \emph{\init}}$ there are no $(k,m),(k,m') \in msg_i[j][t] \cup \mathit{mJ}[t]:m\neq m'$. 
		
	\end{itemize}
	In this case, we say that $c$ is consistent w.r.t. $p_i$. Suppose every system state in $R$ is consistent w.r.t. $p_i$. In this case, we say that $R$ is consistent w.r.t. $p_i$ and Algorithm~\ref{alg:consensus}.
\end{definition}

Note the term active (Definition~\ref{def:consistent}) does not distinguish between the cases in which a node is active due to the occurrence of a transient fault and the invocation of $\mathsf{brbBroadcast}(v)$.

\Subsection{Consistency regaining for Algorithm~\ref{alg:consensus}}

\begin{theorem}[Algorithm~\ref{alg:consensus}'s Convergence]
	\label{thm:recoveryconsensusI}
	Let $R$ be a fair execution of Algorithm~\ref{alg:consensus} in which $p_i\in \sP: i\in \Correct$ is active eventually. 
	%
	%
	The system eventually reaches a state $c \in R$ that starts a consistent execution w.r.t. $p_i$ (Definition~\ref{def:consistent}).
\end{theorem}
\renewcommand{\thmcnt}{\ref{thm:recoveryconsensusI}}
\begin{theoremProof}
	Suppose that $R$'s starting state is not consistent w.r.t. $p_i$. Specifically, suppose that either invariant (brb.i) or (brb.ii) does not hold. \Ie at least one of the if-statement conditions in lines~\ref{ln:consistent1} and~\ref{ln:consistent0} holds. Since $R$ is fair, eventually $p_i$ takes a step that includes the execution of lines~\ref{ln:consistent1} to~\ref{ln:consistent0}, which assures that $p_i$ becomes consistent with respect to (brb.i) and (brb.ii).
	%
	%
	Observe that once invariant (brb.i) and (brb.ii) hold w.r.t. $p_i$ in $c$, they hold in any state $c' \in R$ that follows $c$, cf. lines~\ref{ln:merge} to~\ref{ln:messageConsistentThen} and ~\ref{ln:initEcho} to~\ref{ln:BRBupon}.
	
	
	Due to the above, the rest of the proof assumes, w.l.o.g., that all correct nodes are consistent w.r.t. $p_i$, (brb.i), and (brb.ii) in any state of $R$. 
	Let $m$ be a message that in $R$'s starting state resides in a channel between a pair of correct nodes. 
	%
	%
	Recall that $m$ can reside in that channel only for a finite time (Section~\ref{sec:benignFailures}). 
	Thus, by the definition of complete iterations, the system reaches a state in which $m$ does not appear in the communication channels eventually. Thus, (brb.iii) holds eventually, since it is sufficient to consider only messages that were sent during $R$ from nodes in which (brb.i) and (brb.ii) hold.
\end{theoremProof}



\Subsection{Completion of BRB-broadcast}

\begin{theorem}[BRB-Completion-1]
	\label{thm:brbTerminateSimple}
	Let $typ \in \texttt{brbMSG}$ and $R$ be a consistent execution of Algorithm~\ref{alg:consensus} in which $p_i \in \sP$ is active. 
	Eventually, $\forall i,j \in \Correct: \mathsf{brbDeliver}_j( i) \neq \bot$. 
\end{theorem}
\renewcommand{\thmcnt}{\ref{thm:brbTerminateSimple}}
\begin{theoremProof}
	Since $p_i$ is correct, it broadcasts $\mathrm{MSG}(\mathit{brb}J=msg_i[i],\bull)$ infinitely often. By fair communication, every correct $p_j \in \sP$ receives $\mathrm{MSG}(\mathit{brb}J)$ $=m,\bull)$ eventually. Thus, $\forall j \in \Correct: msg_j[i][\init] = \{m\}$ due to line~\ref{ln:messageConsistentIf}. Also, $\forall j \in \Correct: msg_j[j][\echo] \supseteq  \{(i,m)\}$ since node $p_j$ obverses that the if-statement condition in line~\ref{ln:initEcho} holds (for the case of $k_j=i$). Thus, $p_j$ broadcasts $\mathrm{MSG}(\mathit{brb}J=msg_j[j],\bull)$ infinitely often. By fair communication, every correct node $p_\ell \in \sP$ receives $\mathrm{MSG}(\mathit{brb}J,\bull)$ eventually. Thus, $\forall j,\ell \in \Correct: msg_\ell[j][\echo] \supseteq \{(i,m)\}$ (line~\ref{ln:messageConsistentIf}). Since \emsO{$n\mathit{-}t>(n\mathit{+}t)/2$,} node $p_\ell$ observes that $(n\mathit{+}t)/2<|\{ p_x \in  \sP: (i,m) \in msg_\ell[x][\echo]\}|$ holds, \ie the if-statement condition in line~\ref{ln:echoReady0If} holds, and thus, $msg_\ell[\ell][\ready] \supseteq \{(i,m)\}$ holds. 
	Note that, since $t < (n\mathit{+}t) /2$, faulty nodes cannot prevent a correct node from broadcasting $\mathrm{MSG}(\mathit{brb}J=\mathit{m}J,\bull):\mathit{mJ}[\ready]\supseteq \{(i,m)\}$ infinitely often, say, by colluding and sending $\mathrm{MSG}(\mathit{brb}J=\mathit{m}J,\bull):\mathit{mJ}[\ready]\supseteq \{(i,m')\} \land m'\neq m$. By fair communication, every correct $p_y \in \sP$ receives $\mathrm{MSG}(\mathit{brb}J=\mathit{m}J,\bull)$ eventually. Thus, $\forall j,y\in \Correct: msg_y[j][\ready] \supseteq \{(i,m)\}$ holds (line~\ref{ln:messageConsistentIf}). Therefore, whenever $p_y$ invokes $\mathsf{brbDeliver}_y(i)$ (line~\ref{ln:brbDeliver}), the\remove{ if-statement} condition $\exists_m (2t\mathit{+}1) \leq |\{ p_{\ell} \in \sP : {(k_y=i,m) \in msg_y[\ell][\ready]} \}|$ holds, and thus, $m$ is returned.
\end{theoremProof}


\Subsection{Closure of BRB-broadcast}
\label{sec:brbClosure}
The main difference between the completion and the closure proofs is that the latter considers post-recycled starting system states and complete invocation of operations (Definition~\ref{def:well}).

\begin{definition}[Post-recycle system states and complete invocation of operations]
	\label{def:well}
	We say that system state $c$ is \emph{post-recycle}  w.r.t. $p_i \in \sP: i \in \Correct$ if $\forall  j \in \Correct:msg_j[i] = [\emptyset,\ldots,\emptyset]$ holds and no communication channel from $p_i$ to $p_j$ includes $\mathrm{MSG}(\mathit{brb}J\neq[\emptyset,\ldots,\emptyset],\bull)$. Suppose that execution $R$ starts in the post-recycled system state $c$ and $p_i$ invokes $\mathsf{brbBroadcast}_i(v)$ exactly once. In this case, we say that $R$ includes a \emph{complete BRB invocation} w.r.t. $p_i$.
	%
\end{definition}

Note that a post-recycled system state (Definition~\ref{def:well}) is also a consistent one (Definition~\ref{def:consistent}).

\begin{theorem}[BRB closure]
	\label{thm:brbClousre}
	Let $R$ be a post-recycled execution of Algorithm~\ref{alg:consensus} in which all correct nodes are active eventually via the complete invocation of BRB-broadcast. The system demonstrates in $R$ a BRB construction.
\end{theorem}
\renewcommand{\thmcnt}{\ref{thm:brbClousre}}
\begin{theoremProof}
	BRB-Completion-1 holds (Theorem~\ref{thm:brbTerminateSimple}). 
	
	
	\begin{lemma}[BRB-Completion-2]
		\label{thm:brbTerminateSimple2}
		%
		BRB-Completion-2 holds.
	\end{lemma}
	\renewcommand{\lemcnt}{\ref{thm:brbTerminateSimple2}}
	\begin{lemmaProof}
		By line~\ref{ln:brbDeliver}, $p_i$ can BRB-deliver $m$ from $p_j$ only once $\exists_m (2t\mathit{+}1) \leq |\{ p_{\ell} \in \sP:{(k,m) \in msg_i[\ell][\ready]}\}|$ holds. 
		During post-recycled execution, only lines~\ref{ln:echoReady0} to~\ref{ln:echoReady1} and~\ref{ln:messageConsistentThen} can add items to $msg_i[i][\ready]$ and $msg_i[\ell][\ready]$, respectively. 
		Let $\mathrm{MSG}(\mathit{mJ})$ be such that $\mathit{mJ}[\ready]\supseteq \{(j,m)\}$.
		Specifically, line~\ref{ln:messageConsistentThen} adds to $msg_i[\ell][\ready]$ items according to information in $\mathrm{MSG}(\mathit{mJ})$ messages coming from $p_\ell$. This means, that at least $t+1$ distinct and correct nodes broadcast $\mathrm{MSG}(\mathit{mJ})$ infinitely often. By fair communication and line~\ref{ln:messageConsistentThen}, all correct nodes, $p_x$, eventually receive $\mathrm{MSG}(\mathit{mJ})$ from at least $t+1$ distinct nodes and make sure that $msg_x[\ell][\ready]$ includes $(j,m)$. Also, by line~\ref{ln:echoReady1}, we know that $msg_x[x][\ready] \supseteq \{(j,m)\}$, \ie every correct node broadcast $\mathrm{MSG}(\mathit{mJ})$ infinitely often. By fair communication and line~\ref{ln:messageConsistentThen}, all correct nodes, $p_x$, receive $\mathrm{MSG}(\mathit{mJ})$ from at least $2t+1$ distinct nodes eventually, because there are at least $n\mathit{-}t \geq 2t\mathit{+}1$ correct nodes. This implies that $\exists_m (2t\mathit{+}1) \leq |\{ p_{\ell} \in \sP:{(k,m) \in msg_i[\ell][\ready]}\}|$ holds (due to line~\ref{ln:messageConsistentThen}). Hence, $\forall i \in \Correct: \mathsf{brbDeliver}_i(j) \notin \{\bot,\blitza\}$. 
	\end{lemmaProof}

	\begin{lemma}
		\label{thm:brbIntegrity}
		The BRB-integrity property holds.
	\end{lemma}	
	\renewcommand{\lemcnt}{\ref{thm:brbIntegrity}}
	\begin{lemmaProof}
		Suppose $\mathsf{brbDeliver}(k)=m\neq \bot$ holds in $c \in R$. Also, (towards a contradiction) $\mathsf{brbDeliver}(k)=m'\notin  \{\bot,m\}$ holds in $c' \in R$, where $c'$ appears after $c$ in $R$. \Ie $\exists_m (2t\mathit{+}1) \leq |\{ p_{\ell} \in \sP:{(k,m) \in msg_i[\ell][\ready]}\}|$ in $c$ and $\exists_{m'} (2t\mathit{+}1) \leq |\{ p_{\ell} \in \sP:{(k,m') \in msg_i[\ell][\ready]}\}|$ in $c'$. For any ${i,j,k \in \Correct}$ and any ${typ \in \texttt{brbMSG}}$ it holds that $ (k,m),(k,m') \in msg_i[j][\ready]$ (since $R$ is post-recycle, and thus, consistent). Thus, $m=m'$, cf. invariant (brb.iii). Also, observe from the code of Algorithm~\ref{alg:consensus} that no element is removed from any entry $msg[][][]$ during consistent executions. This means that $msg_i[\ell][\ready]$ includes both $(k,m)$ and $(k,m')$ in $c'$. However, this contradicts the fact that $c'$ is consistent. Thus, $c' \in R$ cannot exist and BRB-integrity holds. 
	\end{lemmaProof}
	
	\begin{lemma}[BRB-validity]
		\label{thm:brbValidity}
		BRB-validity holds.
	\end{lemma}
	\renewcommand{\lemcnt}{\ref{thm:brbValidity}}
	\begin{lemmaProof}
		Let $p_i,p_j :i,j\in \Correct$. Suppose that $p_j$ BRB-delivers message $m$ from $p_i$. The proof needs to show that $p_i$ BRB-broadcasts $m$. In other words, suppose that the adversary, who can capture up to $t$ (Byzantine) nodes, sends the ``fake'' messages of $msg_j[j][\echo] \supseteq  \{(i,m)\}$ or $msg_j[j][\ready] \supseteq  \{(i,m)\}$, but $p_i$, who is correct, never invoked $\mathsf{brbBroadcast}(m)$. In this case, our proof shows that no correct node BRB-delivers $\langle i,m \rangle$. This is because there are at most $t$ nodes that can broadcast ``fake'' messages. Thus, $\mathsf{brbDeliver}(k)$ (line~\ref{ln:brbDeliver}) cannot deliver $\langle i,m \rangle$ since $t < 2t \mathit{+} 1$, which means that the if-statement condition $\exists_m (2t\mathit{+}1) \leq |\{ p_{\ell} \in \sP : {(k,m) \in msg_i[\ell][\ready]} \}|$ cannot be satisfied.
	\end{lemmaProof}

	\begin{lemma}[BRB-no-duplicity]
		\label{thm:brbDuplicity}
		Suppose $p_i, p_j: i,j \in \Correct$, BRB broadcast $\mathrm{MSG}(\mathit{mJ}):\mathit{mJ}[\ready]\supseteq \{(k,m)\}$, and respect., $\mathrm{MSG}(\mathit{mJ}):\mathit{mJ}[\ready]\supseteq \{(k,m')\}$. We have $m = m'$.
	\end{lemma}
	\renewcommand{\lemcnt}{\ref{thm:brbDuplicity}}		
	\begin{lemmaProof}
		Since $R$ is post-recycle, there must be a step in $R$ in which the element $(k,\bull)$ is added to $msg_x[x][\ready]$ for the first time during $R$, where $p_x \in \{p_i,p_j\}$.	The correctness proof considers the following two cases.
		
		
		
		$\bullet$ \textbf{Both $p_i$ and $p_j$ add $(k,\bull)$ due to line~\ref{ln:echoReady0}.~~} Suppose, towards a contradiction, that $m\neq m'$. Since the if-statement condition in line~\ref{ln:echoReady0} holds for both $p_i$ and $p_j$, we know that $\exists_{m} (n\mathit{+}t)/2<|\{ p_{\ell} \in  \sP: (k,m) \in msg_i[\ell][\echo]\}|$ and $\exists_{m'} (n\mathit{+}t)/2<|\{ p_{\ell} \in  \sP: (k,m') \in msg_j[\ell][\echo]\}|$ hold. Since $R$ is post-recycle, this can only happen if $p_i$ and $p_j$ received $\mathrm{MSG}(\mathit{mJ}):\mathit{mJ}[\echo]\supseteq \{(k,m)\}$, and respect., $\mathrm{MSG}(\mathit{mJ}):\mathit{mJ}[\echo]\supseteq \{(k,m')\}$ from $(n\mathit{+}t)/2$ distinct nodes. Note that $\exists p_x \in Q_1 \cap Q_2:x \in \Correct$, where $Q_1,Q_2 \subseteq \sP:|Q_1|,|Q_2| \geq 1\mathit{+}(n\mathit{+}t)/2$ (as in~\cite{DBLP:books/sp/Raynal18}, item (c) of Lemma 3). But, any correct node, $p_x$, has at most one element in $msg_x[\ell][\echo]$ (line~\ref{ln:initEcho}) during $R$. Thus, $m = m'$, which contradicts the case assumption.	
		
		
		$\bullet$ \textbf{There is $p_x \in \{p_i,p_j\}$ that adds $(k,\bull)$ due to line~\ref{ln:echoReady1}.~~} \Ie $\exists_{m''} (t\mathit{+}1) \leq |\{ p_{\ell} \in  \sP: (k,m'') \in msg[\ell][\emph{\ready}]\}| \land m'' \in \{m,m'\}$. Since there are at most $t$ faulty nodes, $p_x$ received $\mathrm{MSG}(\mathit{mJ}):\mathit{mJ}[\ready]\supseteq \{(k,m'')\}$ from at least one correct node, say $p_{x_1}$, which received $\mathrm{MSG}(\mathit{mJ}):\mathit{mJ}[\ready]\supseteq \{(k,m'')\}$ from $p_{x_2}$, and so on. This chain cannot be longer than $n$ and it must be originated by the previous case in which $(k,\bull)$ is added due to line~\ref{ln:echoReady0}. Thus, $m = m'$.
	\end{lemmaProof}
\end{theoremProof}


\Section{Self-stabilizing Recycling in Time-free Message-passing Systems} 
\label{sec:sysFree}
%
%
Before proposing our self-stabilizing BFT algorithm for BRB instance recycling (Section~\ref{sec:sysFreeSystem}), we study a non-crash-tolerant yet self-stabilizing recycling algorithm for time-free systems. Namely, as steppingstones towards a solution for $\mathsf{BAMP_{n,t}[FC,t < n/3, BML, \diamondsuit P_{mute}]}$, we present the \emph{independent round counter} (IRC) task and implement $\mathsf{txAvailable}()$ and $\mathsf{rxAvailable}(k)$ (Figure~\ref{fig:BrbIRC} and Algorithm~\ref{alg:consensus}). 

\emsO{When non-self-stabilizing node-failure-free systems are considered, the operation $\mathsf{txAvailable}()$ and the operation $\mathsf{rxAvailable}(k)$ can be implemented using prevailing mechanisms for automatic repeat request (ARQ), which uses unbounded counters. These mechanisms are often used for guaranteeing reliable communications by letting the sender collect acknowledgments from all receivers. Each message is associated with a unique message number, which the sender obtains by adding one to the previous message number after all acknowledgments arrived. From that point in time, the previous message number is obsolete and can be recycled. For the case of self-stabilizing node-failure-free systems, the challenge is to deal with integer overflow events. Specifically, when an algorithm considers the counters to be unbounded but the studied system has bounded memory, transient faults can trigger integer overflow events. The solution presented here shows how to overcome this challenge via our recycling technique and a mild synchrony assumption.}



\Subsection{Independent Round Counters (IRCs)} 
\label{sec:prbDef}
We consider $n$ independent counters, such that each counter, $cnt_i$, can be incremented only by a unique node, $p_i \in \sP$, via the innovation of the $\mathsf{increment}_i()$ operation, which returns the new round number or $\bot$ when the invocation is (temporarily) disabled. Suppose $p_i,p_j \in \sP$ are correct. Every node $p_j \in \sP$ can fetch $cnt_i$'s value via the invocation of the $\mathsf{fetch}_j(i)$ operation, which returns the most recent and non-fetched $cnt_i$'s value or $\bot$ when such value is currently unavailable. We define the \emph{Independent Round Counters} (IRCs) task using the following requirements.  

\begin{itemize}
	\item \textbf{IRC-validity.~~} Suppose $p_j$ IRC-fetches $s$ from $cnt_i$. Then, $p_i$ had IRC-incremented $cnt_i$ to $s$.

	\item \textbf{IRC-integrity-1.~~} Let $S_{i,j}=(s_0,\ldots,s_{x}):x<B$ be a sequence of $p_i$'s round numbers that $p_j$ fetched---we are only interested in $B$ most recent ones, where $B$ is a predefined constant. It holds that $\forall s_y\in S_{i,j}: y<B-1 \implies s_y+1 \bmod B=s_{y+1}$. In other words, no correct node IRC-fetches a value more than once from the counter of any other correct node (considering the $B$ most recent IRC-fetches).

	\item \textbf{IRC-integrity-2.~~} Correct nodes that IRC-fetch numbers from $cnt_i$ do so in the order in which $cnt_i$ was incremented (considering the $B$ most recent IRC-fetches).

	\item \textbf{IRC-preemption.~~} Suppose $p_i$ IRC-increments $cnt_i$ to $s$. IRC-increment is (temporarily) disabled until all correct nodes have fetched $s$ from $p_i$'s counter.

	\item \textbf{IRC-completion.~~} Suppose all correct nodes, $p_j$, IRC-fetch $p_i$'s counter infinity often. Node $p_i$'s IRC-increment is enabled infinity often.  
\end{itemize}


Note that any algorithm that solve the IRC task can implement the interface functions $\mathsf{txAvailable}()$ and $\mathsf{rxAvailable}(k)$ by returning $\mathsf{increment}() \neq \bot$ and $\mathsf{fetch}(k) \neq \bot$, respectively. 



\Subsection{\emsO{Time-free system settings for $\mathsf{AMP_{n}[FC, BML]}$}}
%
%
\label{sec:SelfIRC}
\emsO{The IRC solution proposed in this section requires time-free system settings, which we define by revising $\mathsf{BAMP_{n,t}[FC,t < n/3]}$ into $\mathsf{AMP_{n}[FC, BML]}$. The latter model does not consider node failures but includes Assumption~\ref{def:synch}, as we explain next.}  
%

Consider a scenario in which, due to a transient fault, $p_i$'s copy of its round counter is smaller than $p_j$'s copy of $p_i$'s counter, say, by $x \in \mathbb{Z}^+$, thus node $p_i$ will have to complete $x$ rounds before $p_j$ could IRC-fetch a non-$\bot$ value. The proposed IRC algorithm overcomes is challenge by following Assumption~\ref{def:synch}.   

\begin{assumption}[Bounded message lifetime, BML]
	\label{def:synch}
	Let $R$ be an execution in which there is a correct node $p_i \in \sP$ that repeatedly broadcasts the protocol messages and completes an unbounded number of round-trips with every correct node, $p_j \in \sP$, in the system. Suppose that $p_j$ receives message $m(s)$ from $p_i$ 
	%
	%
	immediately before system state $c \in R$, where $s \in \mathbb{Z}^+$ is the round number. We assume that $cur_i[i]-s \leq \lambda$ in $c$, where $\lambda \in \mathbb{Z}^+:\capacity < \lambda<B/6$ is a known upper-bound, $\capacity$ is defined in Section~\ref{sec:sys}, and $B$ is defined by line~\ref{ln:cost}.
\end{assumption}

\begin{algorithm}[t!]
	\begin{\algSize}
		\smallskip

		\textbf{constants:}			
		
		\label{ln:cost} $B$: a predefined bound on the integer size, say, $2^{64}-1$.
		
		\smallskip
		
		
		\smallskip
		
		\textbf{variables:}			
		
		%
		
		\label{ln:varSeq} $cur[\sP],nxt[\sP]=[[-1, -1],\ldots,[-1, -1]]$: a pair of round numbers---one pair per system node, where $cur[i]$ is $p_i$'s current round number and $nxt[i]$ is the next one. Also, $cur[j]$ and $nxt[j]$ store the most recently received, and respectively, delivered round number from $p_j$\;
		
		\label{ln:varLbl} $lbl[\sP]=[0,\ldots, 0]$: labels corresponding to $cur[i]$, where $lbl[j]$ stores the most recently received label from $p_j$.
		
		%
		%
		
		\smallskip
		
		\textbf{required interfaces:}  	
		$\mathsf{recycle}(k)$, $\trusted()$, \fbox{$\mdReset()$, $\mdCnt(j)$\label{ln:Freset};}
		
		\smallskip
		
		\textbf{provided interface:}  	
		
		$\mathsf{txAvailable}()$ \label{ln:Tstart} \textbf{do} \{\Return{$\mathsf{increment}() \neq \bot$}\}
		
		$\mathsf{rxAvailable}(k)$ \label{ln:Rstart} \textbf{do} \{\Return{$\mathsf{fetch}(k) \neq \bot$}\}
		
		\smallskip
		
		\textbf{macro:} $\mathsf{behind}(d,s,c)$ \label{ln:behind}\textbf{do} \{\Return{$s \in \{x \bmod B: x\in \{c-d\lambda,\ldots,c\}\}$}\}  	

		\smallskip
		
		\textbf{operation} $\mathsf{increment}()$ \label{ln:increment}\Begin{
						\tcc{$\trusted()=\sP$ for the non-crash-tolerant version}
			\If{$cur[i] = -1  \lor \exists j \in \trusted(): \mathit{lbl}[j] \leq 2(\capacity+1)$\label{ln:incrementIf}}{\Return{$\bot$}\label{ln:incrementThen}}
			\lElse{\fbox{$\mdReset()$;} $cur[i]\gets cur[i]+1\bmod B$; $\mathsf{recycle}(j)$; \remove{\label{ln:BcurPlusOne}}\Return{$cur[i]$}}\label{ln:incrementElse}}


		\smallskip

		\textbf{operation} $\mathsf{fetch}(k)$ \label{ln:fetch}\Begin{
			\lIf{$\mathsf{behind}(1,cur[k],nxt[k])$\label{ln:fetchIf}}{\Return{$\bot$}\label{ln:fetchThen}}\lElse{\{$nxt[k] \gets cur[k]$; \Return{$nxt[k]$}\label{ln:fetchElse}\}}
		}

		
		\smallskip
		
		\textbf{operation} $\mathsf{txMSG}(j)$ \label{ln:Ftx}\{\Return{$(\true, cur[i],\mathit{lbl}[j])$}\}
		
		\smallskip
		
		
		\textbf{operation} $\mathsf{rxMSG}(\mathit{brb}J,\mathit{irc}J=(\mathit{a}J, \mathit{s}J, \mathit{{\ell}J}),j)$ \label{ln:Frx}\Begin{
			
			\If{$\neg \mathit{aJ} \land \mathsf{behind}(2,cur[i], \mathit{sJ}) \land \mathit{lbl}[j] = \mathit{\ell}J$\label{ln:FrxIf}}{\{\fbox{$\mdCnt(j)$;} $\mathit{lbl}[j] \gets \min \{B, \mathit{\ell}J+1\}$; \Return\}\label{ln:FrxThen}}
			
			

			\If{$\neg \mathsf{behind}(1,\mathit{sJ},cur[i])$\label{ln:rxIfSeq}}{\{$cur[j] \gets \mathit{sJ}$\label{ln:rxThenSeq}; $\mathsf{recycle}(j)$\}}

			%
			%

			$\mathbf{send}~ \mathrm{MSG}(\false, nxt[j], \mathit{{\ell}J})$ \textbf{to} $p_j$\label{ln:rxIfSend}\;
		}

		\smallskip
		
		\textbf{do forever} 		
		\textbf{broadcast} $\mathrm{MSG}(\mathit{brb}I=msg[i], \mathit{irc}I=\mathsf{txMSG}())$;\label{ln:ircBroadcast}
		
		
		\smallskip
		
		
		\textbf{upon} $\mathrm{MSG}(\mathit{brb}J, \mathit{irc}J)$\label{ln:IRCupon} \textbf{arrival from} $p_j$  \Begin{$\mathit{mrg}(\mathit{brb}J,j)$\; $\mathsf{rxMSG}(\mathit{brb}J,\mathit{irc}J,j)$\;}
		
		\smallskip

		\caption{\label{alg:IRCnoCrach}time-free IRC; code for $p_i$}
	\end{\algSize}
\end{algorithm}

\Subsection{Self-stabilizing IRC for \emsO{$\mathsf{AMP_{n}[FC, BML]}$}} 
\label{sec:free}
Algorithm~\ref{alg:IRCnoCrach} presents a self-stabilizing solution for crash-free message-passing systems. \Ie it assumes that all nodes are correct. Algorithm~\ref{alg:IRCnoCrach} makes sure that any node that had IRC-incremented its round counter defers any further IRC-increments until all nodes have acknowledged the latest IRC-increment. Note that the line numbers of Algorithm~\ref{alg:IRCnoCrach} continue the ones of Algorithm~\ref{alg:consensus}. Also, the \fbox{boxed} code in lines~\ref{ln:incrementElse} and~\ref{ln:FrxThen} are irrelevant to the IRC solution studied in this section.
We remind that the implementation of interface function $\mathsf{recycle}()$ (line~\ref{ln:Freset}) is provided by Algorithm~\ref{alg:consensus}, line~\ref{ln:reset}. Also, for this section, let us assume that $\trusted_i()=\sP$. 

%
%

\Subsubsection{Constants and variables} 
All integers used by Algorithm~\ref{alg:IRCnoCrach} have a maximum value, which we denote by $B$ (line~\ref{ln:cost}) and require to be large, say, $2^{64}-1$. The arrays $cur[]$ and $nxt[]$ (line~\ref{ln:varSeq}) store a pair of round numbers. The entry $cur[i]$ is $p_i$'s current round number and $nxt[i]$ is the next one. Also, $cur[j]$ and $nxt[j]$ store the most recently received, and respectively, delivered round numbers from $p_j$. 
%
%
The array $lbl[]$ holds labels that correspond to the number in $cur[i]$, where $lbl[j]$ is the most recently received label from $p_j$ (line~\ref{ln:varLbl}).
%

\Subsubsection{The $\mathsf{increment}()$ operation}			
This operation allows the caller to IRC-increment the value of its round number modulo $B$. It also returns the new round number. However, if the previous invocation has not finished, the operation is disabled and the $\bot$ value is returned. Line~\ref{ln:incrementIf} tests whether the round number is ready to be incremented. In detail, recall that in this section, we assume $\trusted_i()=\sP$. Now line~\ref{ln:incrementIf} checks whether this is the first round, \ie a round number of $-1$, or the previous round has finished, \ie the labels indicate that every node has completed at least $2(\capacity+1)$ round trip. By exchanging at least $2(\capacity+1)$ labels, the proposed solution overcomes packet loss and duplication over non-FIFO channels, see~\cite{DBLP:conf/sss/DolevHSS12} for a more efficient variation on this technique. 


\Subsubsection{The $\mathsf{fetch}(k)$ operation}			
This operation returns, exactly once, the most recently received round number. Line~\ref{ln:fetchIf} tests whether a new round number has arrived. If this is not the case, then $\bot$ is returned. Otherwise, the value of the new round number is returned (line~\ref{ln:fetchElse}). In detail, due to Assumption~\ref{def:synch}, immediately after the arrival of message $m(s)$ to $p_j$ from $p_i$, the fact that $s \notin \{x \bmod B: x\in \{c-1\lambda,\ldots,c\}\}$ holds implies that $s$ is newer than $cur_j[i]$. Thus, $p_i$ can use $\mathsf{behind}_i(1,cur_i[k],nxt_i[k])$ (line~\ref{ln:fetchIf}) for testing the freshness of the round number stored in $cur_i[k]$ w.r.t. $nxt_i[k]$. If case the number is indeed fresh, $\mathsf{fetch}_i()$ updates $nxt_i[k]$ with the returned round number.

\Subsubsection{The $\mathsf{txMSG}()$ and $\mathsf{rxMSG}()$ operations}			
The operations $\mathsf{txMSG}()$ and $\mathsf{rxMSG}()$ let the sender, and respectively the receiver, process messages. Algorithm~\ref{alg:IRCnoCrach} sends via the message $MSG()$ two fields: $\mathit{brb}J$ and $\mathit{irc}J$, where the field $\mathit{brb}J$ is related to Algorithm~\ref{alg:consensus}. Recall that when a message arrives from $p_j$, the receiving-side adds the suffix $J$ to the field name, \ie $\mathit{brb}J$ and $\mathit{irc}J$. The field $\mathit{irc}J$ is composed of the fields $ack$, which indicates whether acknowledge is required, $seq$, which is the sender's round number, and $lbl$, which, during legal executions, is the corresponding label to $seq$ that the sender uses for the receiver. 

The operation $\mathsf{txMSG}()$ is used when the sender transmits a message (line~\ref{ln:Ftx}). It specifies that acknowledgment is required, \ie $ack=\true$ as well as includes the sender's current round number, \ie $cur[i]$, and the corresponding label that the sender uses for the receiver $p_j \in \sP$, \ie $\mathit{lbl}[j]$.


The operation $\mathsf{rxMSG}()$ processes messages arriving either to the sender or the receiver. On the sender-side, when an acknowledgment arrives from receiver, $p_j$, the sender checks whether the arriving message has fresh round number and label (line~\ref{ln:FrxIf}). In this case, the label is incremented in order to indicate that at least one round trip was completed. In detail, $p_i$ uses $\mathsf{behind}_i(2,cur_i[j], \mathit{sJ})$ for testing whether the arriving round number, $\mathit{sJ}$, is fresh by asking whether $\mathit{sJ}$ is not a member of the set $\{cur_j[i]-2\lambda,\ldots,cur_j[i]\}$, see Assumption~\ref{def:synch}. As we will see in the next paragraph, there is a need to take into account the receiver's test (line~\ref{ln:rxIfSeq}), which can cause a non-fresh value to be a member of the set $\{x \bmod B: x\in \{c-2\lambda,\ldots,c\}\}$, but not the set $\{x \bmod B: x\in \{c-\lambda,\ldots,c\}\}$.

On the receiver-side, $p_i$ uses $\mathsf{behind}_i(1, \mathit{sJ}, cur_i[j])$ to test whether a new round number arrived, \ie testing whether the arriving number, $\mathit{sJ}$, is a member of $\{cur_j[i]-2\lambda,\ldots,cur_j[i]\}$. In this case, the local round number is updated (line~\ref{ln:rxThenSeq}) and the interface function $\mathsf{recycle}_i(j)$ is called (line~\ref{ln:reset}).
%
%
%
Note that whenever the receiver gets a message, it replies (line~\ref{ln:rxIfSend}). That acknowledgment specifies that no further replies are required, \ie $ack=\false$, as well as the most recently delivered round number, \ie $nxt[i]$, and label, $\mathit{{\ell}J}$.     

\Subsubsection{The do forever loop and message arrival}			
Note that the processing of messages (for sending and receiving) is along the lines of Algorithm~\ref{alg:consensus}. The do forever loop broadcasts the message $\mathrm{MSG}()$ to every node in the system (line~\ref{ln:ircBroadcast}). The operation $\mathsf{txMSG}()$ is used for setting the value of the $\mathit{irc}J$ field. Upon message arrival, the receiver passes the arriving values to $\mathsf{rxMSG}()$ for processing (line~\ref{ln:IRCupon}).

\Subsection{Correctness of Algorithm~\ref{alg:IRCnoCrach}}			
The proof is implied by Theorem~\ref{thm:nonSelfIRCnoCrash}.
\begin{theorem}
	\label{thm:nonSelfIRCnoCrash}
	Let $R$ be an Algorithm~\ref{alg:IRCnoCrach}'s execution and $i \in \Correct$. Suppose all correct nodes, $p_j$, IRC-fetch $p_i$'s counter infinity often and $p_i$ invokes IRC-increment infinity often. $R$ eventually demonstrates an IRC construction (Section~\ref{sec:prbDef}). 
\end{theorem}
\renewcommand{\thmcnt}{\ref{thm:nonSelfIRCnoCrash}}
\begin{theoremProof}
	\begin{lemma}
		\label{thm:nonSelfIRCnoCrashX}
		%
		The system demonstrates IRC-completion in $R$ (Section~\ref{sec:prbDef}). 
	\end{lemma}
	\renewcommand{\lemcnt}{\ref{thm:nonSelfIRCnoCrashX}}
	\begin{lemmaProof}
		Recall that $p_i$'s IRC-increment is enabled whenever $\mathsf{increment}_i()$ can return a non-$\bot$ value (line~\ref{ln:incrementIf}), where $p_i$ is a correct node. Also, the return of a non-$\bot$ value implies that the value of $cur_i[i]$ changes (line~\ref{ln:incrementElse}). Thus, towards a contradiction, assume $cur_i[i]=s\geq0$ holds in every system state of $R$. The following arguments show the contradiction by demonstrating that, for any correct node $p_j$, the if-statement condition in line~\ref{ln:FrxIf} holds eventually for any $p_j$'s reply $\mathrm{MSG}(\bull, (\false, \bullet))$ arriving to $p_i$.  Note that once $p_i$ executes line~\ref{ln:FrxThen} at least once for every $p_j$, $\mathsf{increment}_i()$  is enabled since the if-statement in line~\ref{ln:incrementIf} does not hold.   
		To show that the predicate $\mathsf{behind}_i(2,cur_i[i], \mathit{sJ})$ holds, we note that $p_i$ is a correct node that broadcasts $\mathrm{MSG}(\bull, (\true, cur_i[i]=s, \bull))$ infinitely often (line~\ref{ln:ircBroadcast}). Thus, every correct node $p_j$ receives $\mathrm{MSG}(\bull, (\true, cur_i[i]=s, \bull))$ infinitely often (due to the communication fairness assumption). In the system state that immediately follows this message arrival (line~\ref{ln:IRCupon}), the if-statement condition in line~\ref{ln:rxThenSeq} holds, \ie $\mathsf{behind}_j(1,s, cur_j[i])$ holds. By the assumption that $p_j$ invokes $\mathsf{fetch}_j(i)$ infinitely often, we know that the if-statement condition in line~\ref{ln:fetchIf} eventually holds. \Ie $\mathsf{behind}_j(1,cur_j[i],s')$ and $\mathsf{behind}_i(2,s,s')$ hold, where $nxt_j[i]=s'$ is the value used when $p_j$ sends $\mathrm{MSG}(\bull, (\false, nxt_j[i]=s', \bull))$ to $p_i$. Thus, once $\mathrm{MSG}(\bull, (\false, \mathit{sJ}=s', \bull))$ arrives to $p_i$ the predicate $\mathsf{behind}_i(2,cur_i[i]=s, \mathit{sJ}=s')$ holds. The proof of $\mathit{lbl}_i[j] = \mathit{\ell}J$ is by fixing the value of $lbl_i[j]=\ell$ and observing that the values of the messages $\mathrm{MSG}(\bull, (\true, cur_i[i]=s, \ell))$ from $p_i$ to $p_j$ and $\mathrm{MSG}(\bull, (\false, cur_i[i]=s, \ell))$ from $p_j$ to $p_i$. 
	\end{lemmaProof}
	
	\begin{lemma}
		\label{thm:nonSelfIRCnoCrashL}
		Eventually, the system demonstrates IRC-validity in $R$ (Section~\ref{sec:prbDef}). 
	\end{lemma}
	\renewcommand{\lemcnt}{\ref{thm:nonSelfIRCnoCrashL}}
	\begin{lemmaProof}
		W.l.o.g. suppose $R$ is the suffix of execution $R'=R'' \circ R$, such that $\mathsf{increment}_i()$ returns a non-$\bot$ value more than $2(\lambda+1)$ times during $R''$. We show that IRC-validity holds in $R$.  \Ie suppose a correct node $p_j$ IRC-fetches $s$ in step $a_j \in R$ from $p_i$'s counter. We show that $p_i$ had IRC-incremented $cnt_i$ to $s$ in step $a_i$ that appears in $R$ before $a_j$. Suppose, towards a contradiction, $\nexists a_i \in R'$, yet $a_j$ returns $s \neq \bot$ from $\mathsf{fetch}_j(i)$ when executing line~\ref{ln:fetchElse}. 

The starting system state of $R'$, the $\mathit{irc}J.\mathit{{\ell}J}$ field of the messages in the communication channels between $p_i$ and $p_j$ and the variables $lbl_i[j]$ and $\mathit{{\ell}J}$ include at most $2(\capacity+1)$ different labels. Since $p_i$ does not change $cnt_i[i]$ before counting the reception of more than $2(\capacity+1)$ labels (line~\ref{ln:incrementIf}), during the period in which $\mathsf{increment}_i()$ returns non-$\bot$ values at least twice, the messages in the channels between $p_i$ and $p_j$ and $p_j$'s variables do not include values that have not changed since the starting system state of $R'$. 
Thus, $p_i$ completes an unbounded number of round-trips with $p_j$ with values that $p_i$ indeed sent.
%

Recall that $a_j$ returns $s\neq\bot$ from $\mathsf{fetch}_j(i)$ when executing line~\ref{ln:fetchElse}. This can only happen when $cur_j[i]=s\neq \bot$. Node $p_j$ assigns $s$ to $cur_j[i]:i\neq j$ only in line~\ref{ln:rxThenSeq} when it processes a message coming from the sender $p_i$. However, $p_i$ can assign $s$ to $cur_i[i]$ only at line~\ref{ln:incrementElse}, \ie $a_i$ exists. We clarify the last argument: by $\mathsf{behind}()$'s definition (line~\ref{ln:behind}), there could be at most $2\lambda$ consecutive times in which $\mathsf{behind}(2,\bullet)$ holds in the if-statement condition in line~\ref{ln:FrxIf} and yet, $cur_j[i]$ has not changed while $cur_i[i]$ has.
	\end{lemmaProof}
	
	\begin{lemma}
		\label{thm:nonSelfIRCnoCrashM}
		Eventually, the system demonstrates in $R$ an IRC construction (Section~\ref{sec:prbDef}). 
	\end{lemma}
	\renewcommand{\lemcnt}{\ref{thm:nonSelfIRCnoCrashM}}
	\begin{lemmaProof}
		Recall that lemmas~\ref{thm:nonSelfIRCnoCrashX} and~\ref{thm:nonSelfIRCnoCrashL} demonstrate IRC-completion and IRC-validity. Thus, w.l.o.g. we can assume that IRC-validity holds throughout $R$. 
		
		\noindent \textbf{IRC-preemption.~~} Suppose that there is a correct node, $p_k \in \sP$, that does not IRC-fetch $s$ from $p_i$'s round counter during $R$ even after $a_i$ (in which $p_i$ IRC-increment $cnt_i$ to $s$). Also, let $a'_i$ be a step that appears in $R$ after $a_i$ and includes an IRC-increment invocation by $p_i$. We show that $a_i$'s invocation returns $\bot$, \ie $a_i$'s invocation is disabled.
		
		By the code of Algorithm~\ref{alg:IRCnoCrach}, the fact that there is no step in $R$ in which $\mathsf{fetch}_k(i)$ returns $s$ implies that $nxt_k[i] \neq s$ holds in any system state during $R$ (line~\ref{ln:fetchElse}). Therefore, $p_k$ does not send $\mathrm{MSG}((\false, nxt[j]=s, \bull), \bull)$ to $p_i$. This means that, as long as $seq_i[i]=s$, it holds that $\mathit{lbl}[k]=0$. Also, as long as $seq_i[i]=s$, whenever $p_i$ invokes $\mathsf{increment}_i()$, the if-statement condition in line~\ref{ln:incrementIf} holds. Thus,  $\mathsf{increment}_i()$ returns $\bot$ in step $a'_i$.   
		
		\noindent \textbf{IRC-integrity-1.~~} Lines~\ref{ln:fetchIf} to~\ref{ln:fetchElse} implies that no correct node, $p_i$, can IRC-fetch the same value twice from the counter of the same node, say, $p_j$.
		
		\noindent \textbf{IRC-integrity-2.~~} Suppose $p_j$ IRC-fetches $s'$ from $p_i$'s counter in step $a'_j$ that appears in $R$ after $a_j$ (in which $p_j$ IRC-fetches $s'$). We show that $p_i$ IRC-incremented $cnt_i$ to $s$ and then to $s'$. Step $a'_i$ appears in $R$ after $a_j$ (IRC-preemption) and $a_j$ after $a_i$ (IRC-validity and line~\ref{ln:incrementElse}). Note that $s\neq s'$, (IRC-integrity-1) \ie $a_i \neq a'_i$. By line~\ref{ln:incrementElse}, $s$ was IRC-incremented before $s'$ when considering the $B$ IRC-increments preceding $a'_i$.	\end{lemmaProof}
\end{theoremProof}



\Section{Self-stabilizing Byzantine-Tolerance IRC via Muteness Detection}
\label{sec:sysFreeSystem}
Algorithm~\ref{alg:IRCnoCrach} presents our self-stabilizing BFT recycling mechanism for \emsO{$\mathsf{BAMP_{n,t}[FC,t < n/3, BML, \diamondsuit P_{mute}]}$, which we obtain by enriching $\mathsf{BAMP_{n,t}[FC,t < n/3, BML]}$ with $\diamondsuit P_{mute}$, which is a detector for muteness failures that we define in Section~\ref{sec:sysMute}.}   

The proposed solution includes the \fbox{boxed} code lines. 
%
%
Algorithm~\ref{alg:IRCnoCrach} lets $p_i$ restart the local state of the muteness detector via a call to $\mdReset_i()$ (line~\ref{ln:incrementElse}). The algorithm uses $\mdCnt_i(j)$ (line~\ref{ln:FrxThen}) for taking into account the completion of a round-trip between $p_i$ and $p_j$. The correctness proof shows (Theorem~\ref{thm:MFD}) that this version of the algorithm can consider $\trusted_i() \subseteq \sP$ due to the properties of $\diamondsuit P_{mute}$ (Section~\ref{sec:specMute}). 



\Subsection{Muteness Failures}
\label{sec:sysMute}
Let us consider an algorithm, $Alg$, that attaches a round number, $seq \in \mathbb{Z}^+$, to every message, $m(seq)$ that it sends. Suppose there is a system state $c_{\tau} \in R$ after which $p_j$, stops forever replying to $p_i$'s messages, $m(seq)$, where $p_i,p_j\in \sP$. In this case, we say that $p_j$ is \emph{mute} to $p_i$ with respect to message $m(seq)$. We clarify that a Byzantine node is not mute if it forever sends all the messages required by $Alg$. For the sake of a simple presentation, we assume that the syntax of $m(seq)$ corresponds to the syntax of a message generated by $Alg$ (since, otherwise, the receiver may simply omit messages with syntax errors). Naturally, the data load of those messages can be wrong. Observe that the set of mute nodes also includes all crashed nodes.

\Subsection{Muteness Detection: Specifications of $\diamondsuit P_{mute}$}
\label{sec:specMute}
We deal with mute nodes via the use of the class $\diamondsuit P_{mute}$ of muteness detectors. In the context of self-stabilization, one has to consider the scenario in which the muteness detector suspects a node due to a transient fault. Thus, the muteness detector has to be restarted from time to time. In this work takes the approach in which one restart occurs at the start of a new round.      



\noindent \textbf{Muteness Strong Completeness:~} Eventually, every mute node is forever suspected w.r.t. round number $s$ by every correct node (or the round number changes).

\noindent \textbf{Eventual Strong Accuracy:~} Eventually, the system reaches a state $c_{\tau} \in R$ in which no correct node is suspected.

\Subsection{Muteness Detection: \emsO{our and related solutions in a nutshell}} 
\label{def:relateMute}
In the context of self-stabilizing Byzantine-free \emsO{(crash-prone)} systems, Beauquier and Kekkonen{-}Moneta~\cite{DBLP:journals/ijsysc/BeauquierK97} and Blanchard \etal~\cite{DBLP:conf/netys/BlanchardDBD14} implemented perfect failure detectors, \ie class $P$, by letting node $p_i$ to suspect any node $p_j \in \sP$ whenever $p_i$ was able to complete $\Theta$ round-trips with other nodes in $\sP$ but not with $p_j$, where $\Theta$ is a predefined constant.

Since the studied fault model includes Byzantine failures, we cannot directly borrow earlier proposals, such as the ones in~\cite{DBLP:journals/ijsysc/BeauquierK97,DBLP:conf/netys/BlanchardDBD14}. Consider, for example, a Byzantine node that anticipates the sender's messages and transmits acknowledgments before the arrival of perceptive messages. Using this attack of speculative acknowledgments, the adversity may accelerate the (false) completion round-trips and let the unreliable failure-detector suspect non-faulty nodes.

\emsO{As we explain next, our solution relies on an assumption (Assumption~\ref{def:synchA}), which facilitates the defense against the above attacks that use speculative acknowledgments. Specifically, when testing whether the $\Theta$ threshold has been exceeded, $p_i$ ignores the round-trips that were completed with the top $t$ nodes, say w.l.g. $p_1,\ldots,p_t$, that had the highest number of round-trips with $p_i$. Suppose w.l.g. that nodes $p^{byz}_{n-t},\ldots,p^{byz}_{n-1}$ are captured by the adversary. On the one hand, the adversary aims at letting $p^{byz}_{n-t},\ldots,p^{byz}_{n-1}$ to rapidly complete round trips with $p_i$. While on the other hand, if any of the nodes $p^{byz}_{n-t},\ldots,p^{byz}_{n-1}$ complete round trips with $p_i$ faster than any of the nodes $p_1,\ldots,p_t$ are ignored by $p_i$ when testing whether the $\Theta$ threshold has been exceeded. In other words, any adversarial strategy that lets any of the nodes $p^{byz}_{n-t},\ldots,p^{byz}_{n-1}$ to complete more round trips with $p_i$ than the nodes $p_1,\ldots,p_t$ cannot cause a ``haste'' muteness detection of a correct node.}

\begin{algorithm}[t!]
	\begin{\algSize}
		\smallskip
		
		\textbf{constants:}			
		
		$B$: a predefined bound on the integer size, say, $2^{64}-1$.
		
		\smallskip
		
		\textbf{variables:}			
		
		$\mathit{rt}[\sP\setminus \{p_i\}][\sP\setminus \{p_i\}]$: round trip counters, initially all entries are zero\label{ln:varrtP}\;
		
		\smallskip
		
		\textbf{interface functions:}
		
		$\mdReset()$ \textbf{do} \{$\mathit{rt} \gets [[0,\ldots,0], \ldots, [0,\ldots,0]]$\}\label{ln:varReset}
		\;

		$\mdCnt(j)$  \Begin{
			\ForEach{$p_k \in \sP \setminus \{p_i,p_j\}$}{$\mathit{rt}[k][j] \gets \min \{B, \mathit{rt}[k][j] +1\}$\label{ln:fAckArMid}}
			$\mathit{rt}[j] \gets [0, \ldots, 0]$;}

		$\trusted()$ \label{ln:trusted} \textbf{do return} {$\{p_j \in \sP:\Theta>\sum_{x \in \mathit{withoutTopItems}(t,j) }x\}$ \textbf{where} $\{\mathit{rt}[j][\ell]\}_{p_\ell \in \sP}$ is a multi-set with all the values in $\mathit{rt}[j][]$ and $\mathit{withoutTopItems}(t,j)$ is the same multi-set after the removal of the top $t$ values;}
		
		\smallskip
		
		\caption{\label{alg:evenMutenessDecetor}Class $\diamondsuit P_{mute}$ detector; code for $p_i$}
	\end{\algSize}
\end{algorithm}

\Subsubsection{Muteness Detection: Implementation} 
\label{def:impleMute}
%
%
As shown in Figure~\ref{fig:BrbIRC}, 
Algorithm~\ref{alg:evenMutenessDecetor} does not send independent messages as it merely provides three interface functions  to 
Algorithm~\ref{alg:IRCnoCrach}, \ie $\mdReset()$, $\mdCnt(j)$, and $\trusted()$. The algorithm's state is based on the array $\mathit{rt}[][]$ (line~\ref{ln:varrtP}), which stores the number of round trips that node $p_i$ has completed with $p_j$. Note that $\mathit{rt}[][]$ counts separately the number of round-trips $p_i$ and $p_k$ are able to complete during any period in which $p_i$ and $p_j$ are attempting to complete a single round-trip.

The function $\mdReset()$ (line~\ref{ln:varReset}) nullifies the value of $\mathit{rt}[][]$. We require that, every time $p_i$ has completed with $p_j$, it calls the $\mdCnt(j)$ (line~\ref{ln:fAckArMid}). This function increments, for every $p_k \in \sP\setminus \{p_i,p_j\}$, the counter in $\mathit{rt}_i[k][j]$. Then, $\mdCnt(j)$ assigns zero to every entry in $\mathit{rt}_i[j]$. The function $\trusted()$ returns the set of unsuspected nodes. Its implementation relies on Assumption~\ref{def:synchA}, which answers to the above challenge (Section~\ref{def:relateMute}). As a defense against the above attacks that use speculative acknowledgment, $p_i$ ignores the top $t$ round-trip counters when testing whether the $\Theta$ threshold has been exceeded. The correctness proof of Algorithm~\ref{alg:evenMutenessDecetor} appears in Theorem~\ref{thm:MFD}.

\begin{assumption}
	\label{def:synchA}
	Let $R$ be an execution in which there is a correct node $p_i \in \sP$ that repeatedly broadcasts the protocol message $m(s):s \in \mathbb{Z}^+$ and completes an unbounded number of round-trips of message $m(s)$ with every correct node in the system. Let $\mathit{rt}_{i,c}:\sP\times\sP\rightarrow Z^+$ be a function that maps any pair of nodes $p_j,p_k \in \sP$ with the number of round-trips that $p_i$ has completed with $p_k$ between system states $c' \in R$ and $c \in R$, where $c'$ is the first system state that immediately follows the last time $p_i$ has completed a round-trip with $p_j$, or the start of $R$ (in case $p_i$ has not completed any round trip with $p_j$ between $R$'s start and $c$). Let $\sum_{x \in \mathit{withoutTopItems}_{i,c}(t,j)}x$ be the total number of round trips that $p_i$ has completed until $c$ when excluding the top $t$ values of $\mathit{rt}_{i,c}$  that have completed with $p_i$ the greatest number of round-trips. 
	%
	%
	We assume that if $\Theta \leq  \sum_{x \in \mathit{withoutTopItems}_{i,c}(t,j)}x$ then $p_j$ is mute to $p_i$  w.r.t. $m(s)$, where $\Theta$ is a predefined constant.
\end{assumption}

\begin{theorem}
	\label{thm:MFD}
	Let $R$ be a legal execution of algorithms~\ref{alg:IRCnoCrach} and~\ref{alg:evenMutenessDecetor} that satisfies Assumption~\ref{def:synchA}. The system demonstrates in $R$ a construction of class $\diamondsuit P_{mute}$ muteness detector (Section~\ref{sec:specMute}).
\end{theorem}
\renewcommand{\thmcnt}{\ref{thm:MFD}}
\begin{theoremProof}
	Let us consider the sequence of values of $\mathit{rt}_i[j][k]$ in the different system states $c \in R$. Note that this sequence is defined by the function $\mathit{rt}_{i,c}(k,j)$ (Assumption~\ref{def:synchA}). Thus, by line~\ref{ln:trusted}, we kt,now that $j \in \trusted_i()$ if, and only if, $\Theta \leq  \sum_{x \in \mathit{withoutTopItems}_{i,c}(t,j)}x$. Let $a_i \in R$ be a step in which $p_i$ invokes $\mathsf{increment}_i()$ and thus calls $\mdReset_i()$ (line~\ref{ln:incrementElse}). We demonstrate that the $\diamondsuit P_{mute}$ class properties hold (Section~\ref{sec:specMute}).
	
	\noindent \textit{Muteness strong completeness:} We show that, eventually, every mute node, $p_m \in\sP$, is forever suspected w.r.t. round number $s$ by every correct node (or the round number is not $s$). Suppose that the round number is always $s$. By the proof of Lemma~\ref{thm:nonSelfIRCnoCrashX}, $p_i$ will call $\mdCnt_i(j)$ infinitely often (line~\ref{ln:FrxThen}). \Ie for every correct node $p_k \in \sP \setminus \{p_i,p_j\}$, the value of $\mathit{rt}_i[k][j]$ will reach the upper bound $B$ eventually. Since $B (n/3)>\Theta$, eventually $p_j \notin \trusted_i()$ holds. 
	
	\noindent \textit{Eventual Strong Accuracy:} We show that eventually, the system reaches a state $c_{\tau} \in R$ in which every correct node, $p_{\ell} \in\sP$, appears in $ \trusted_i()$. Since both $p_i$ and $p_{\ell}$ are correct, we know that $p_i$ completes round-trips with $p_{\ell}$ infinitely often. Whenever a round trip is completed, $p_i$ assigns $[0, \ldots, 0]$ to $\mathit{rt}_i[\ell]$ (due to lines~\ref{ln:FrxThen} and~\ref{ln:fAckArMid}) and the condition $\Theta>\sum_{x \in \mathit{withoutTopItems}_i(t,\ell) }x$ (line~\ref{ln:trusted}) hold until the next round trip completion (Assumption~\ref{def:synchA}).
\end{theoremProof}


\Section{Discussion}
\label{sec:con}
To the best of our knowledge, this paper presents the first SSBFT algorithms for IRC and repeated BRB (that follows Definition~\ref{def:prbDef}) for hybrid asynchronous/time-free systems. As in BT, the SSBFT BRB algorithm takes several asynchronous communication rounds of $\bigO(n^2)$ messages per instance whereas the IRC algorithm takes $\bigO(n)$ messages but requires synchrony assumptions. 

The two SSBFT algorithms are integrated via specified interfaces and message piggybacking (Fig.~\ref{fig:BrbIRC}). Thus, our SSBFT repeated BRB solution increases BT's message size only by a constant per BRB, but the number of messages per instance stays similar. The integrated solution can run an unbounded number of (concurrent and independent) BRB instances. The advantage is that the more communication-intensive component, \ie SSBFT BRB, is not associated with any synchrony assumption. Specifically, one can run $\delta$ concurrent BRB instances, where $\delta$ is a parameter for balancing the trade-off between fault recovery time and the number of BRB instances that can be used (before the next $\delta$ concurrent instances can start). 
The above extension mitigates the effect of the fact that, for the repeated BRB problem, muteness detectors are used and mild synchrony assumptions are made in order to circumvent well-known impossibilities, \eg~\cite{DBLP:journals/jacm/FischerLP85}. 
Those additional assumptions are required for the entire integrated solution to work.
To the best of our knowledge, there is no proposal for a weaker set of assumptions for solving the studied problem in a self-stabilizing manner.

We note that the above extension facilitates the implementation of FIFO-ordered delivery SSBFT repeated BRB. Here, each of the $\delta$ instances is associated with a unqiue label $\ell \in \{0,\ldots, \delta-1\}$. The implementation makes sure that no node $p_i$ delivers a BRB message with label $\ell>0$ before all the BRB messages with labels in $\{0,\ldots, \ell-1\}$. (For the case of $\ell=0$, the delivery is unconditional.)

We hope that the proposed solutions, \eg the proposed recycling mechanism and the hybrid composition of time-free/asynchronous system settings, will facilitate new SSBFT building blocks.         

\bigskip

\noindent \textbf{Acknowledgments.~~}
We are grateful for the comments made by anonymous reviewers that helped to improve the presentation of this article.

\Section{Glossary}

For the reader's convenience, we provide the following list of abbreviations.   

\begin{itemize}
	\item 
AMP a fault model for asynchronous message-passing systems.
	\item 
BAMP a fault model for Byzantine asynchronous message-passing systems.
	\item 
BFT the design criteria of Byzantine fault-tolerant.
	\item 
BML a synchrony assumption about bounded message lifetime, $\lambda$. 
	\item 
BRB the problem of Byzantine Reliable Broadcast.
	\item 
BT the studied non-self-stabilizing BFT algorithm by Bracha and Toueg~\cite{DBLP:journals/jacm/BrachaT85,DBLP:conf/podc/BrachaT83}.
	\item 
FC the fault model-related assumption about fair communications.
	\item 
IRC the problem abstraction of independent round counter, which is used for implementing the proposed BRB-instance recycling for repeated BRB.   
	\item 
RB the problem of Reliable Broadcast.
	\item 
SSBFT self-stabilizing Byzantine fault-tolerant.
	\item 
$n$ number of nodes in the system.
	\item 
$t$ an upper bound on the number of faulty nodes.
	\item 
$\capacity$ an upper bound on the number of messages in any give communication channel.
	\item 
$\delta$ a constant of concurrent BRB instances.   
	\item 
$\lambda$ a bound on the BML lifetime. 
	\item 
$\diamondsuit P_{mute}$ a class of mute failure detectors. 
\end{itemize}


\end{document}